 \documentclass[final,5p,times,authoryear]{elsarticle}
\usepackage{amssymb}
\usepackage{lipsum}
\usepackage{graphicx}
\usepackage{subcaption}
\usepackage{amsmath}
\usepackage{natbib}
\usepackage{comment}
\usepackage{float}
\usepackage[graphicx]{realboxes}
\usepackage{pifont}% http://ctan.org/pkg/pifont
\begin{document}

\begin{frontmatter}

%% Title, authors and addresses

%% use the tnoteref command within \title for footnotes;
%% use the tnotetext command for theassociated footnote;
%% use the fnref command within \author or \affiliation for footnotes;
%% use the fntext command for theassociated footnote;
%% use the corref command within \author for corresponding author footnotes;
%% use the cortext command for theassociated footnote;
%% use the ead command for the email address,
%% and the form \ead[url] for the home page:
%% \title{Title\tnoteref{label1}}
%% \tnotetext[label1]{}
%% \author{Name\corref{cor1}\fnref{label2}}
%% \ead{email address}
%% \ead[url]{home page}
%% \fntext[label2]{}
%% \cortext[cor1]{}
%% \affiliation{organization={},
%%            addressline={}, 
%%            city={},
%%            postcode={}, 
%%            state={},
%%            country={}}
%% \fntext[label3]{}

\title{VGTL-net: Amplitude-Independent Machine Learning for PPG through Visibility Graphs and Transfer Learning}

%% use optional labels to link authors explicitly to addresses:
%% \author[label1,label2]{}
%% \affiliation[label1]{organization={},
%%             addressline={},
%%             city={},
%%             postcode={},
%%             state={},
%%             country={}}
%%
%% \affiliation[label2]{organization={},
%%             addressline={},
%%             city={},
%%             postcode={},
%%             state={},
%%             country={}}

%\author[first,corref]{Yuyang Miao}
\author[a]{Yuyang Miao\corref{cor1}}
\ead{ym520@ic.ac.uk}
\cortext[cor1]{Corresponding author: Yuyang Miao}

\author[a]{Harry J. Davies}
\ead{harry.davies14@imperial.ac.uk}

\author[a]{Danilo P. Mandic}
\ead{d.mandic@imperial.ac.uk}

\affiliation[a]{organization={ Department of Electrical and Electronic Engineering, Imperial College London},%Department and Organization
            addressline={Exhibition Rd, South Kensington}, 
            city={London},
            postcode={SW7 2BX}, 
            country={United Kingdom}}

\begin{abstract}
%% Text of abstract
Photoplethysmography (PPG) refers to the measurement of variations in blood volume using light and is a feature of most wearable devices. The PPG signals provide insight into the body's circulatory system and can be employed to extract various bio-features, such as heart rate and vascular ageing. Although several algorithms have been proposed for this purpose, many exhibit limitations, including heavy reliance on human calibration, high signal quality requirements, and a lack of generalisation. In this paper, we introduce a PPG signal processing framework that integrates graph theory and computer vision algorithms, to provide an analysis framework which is amplitude-independent and invariant to affine transformations. It also requires minimal preprocessing, fuses information through RGB channels and exhibits robust generalisation across tasks and datasets. The proposed VGTL-net achieves state-of-the-art performance in the prediction of vascular ageing and demonstrates robust estimation of continuous blood pressure waveforms.
\end{abstract}

%%Graphical abstract
%\begin{graphicalabstract}
%\includegraphics{grabs}
%\end{graphicalabstract}

%%Research highlights
%\begin{highlights}
%\item Research highlight 1
%\item Research highlight 2
%\end{highlights}

\begin{keyword}
%% keywords here, in the form: keyword \sep keyword, up to a maximum of 6 keywords
Photoplethysmography (PPG) \sep Visibility Graph \sep Transfer Learning \sep Graph Theory

%% PACS codes here, in the form: \PACS code \sep code

%% MSC codes here, in the form: \MSC code \sep code
%% or \MSC[2008] code \sep code (2000 is the default)

\end{keyword}

\end{frontmatter}

%\tableofcontents

%% \linenumbers

%% main text

\section{Introduction}
\label{sec:intro}
With the rapid development of technology, the integration of technology into healthcare, denoted as E-health, is reshaping the ways in which the healthcare services are provided and accessed \citep{eysenbach2001health}. E-health promises enhanced efficiency in healthcare delivery, potentially reducing the cost while simultaneously improving the quality of care.

Photoplethysmography (PPG) sensors play a crucial role in e-health and are widely employed in wearable devices due to their non-invasive and portable nature. Through estimation of blood volume at the site of the probe, PPG can be used to estimate heart rate \citep {biagetti2019reduced}, respiration \citep {motin2017ensemble}, blood pressure \citep {long2023bpnet}, blood oxygen levels \citep {davies2020ear} and vascular stiffness, among others. A wide range of applications based on PPG signals include, but are not limited to, cognitive load classification \citep {9851405, parreira2023proof}, blood glucose monitoring \citep {hammour2023ear, 9944662} and Chronic Obstructive Pulmonary Disease (COPD) diagnosis \citep {davies2022wearable}.

Numerous studies have focused on extracting respiratory information from PPG signals \citep {sultan2023continuous, motin2019selection, davies2023rapid}.  
Harry \textit{et al.} \citep {davies2023rapid} applied corr-encoder directly on to the PPG signal and extracted the breathing envelope and estimated the breathing rate on it. Muhammad \textit{et al.} \citep {sultan2023continuous} proposed using multiple features fore breathing rate extraction. Physio, spectral and statistical features were extracted and selected based on frequency analysis, correlation coefficients, mutual information and minimal redundancy maximal relevance (mRMR). Finally the breathing rate was determined using a Neural Network based on the selected features. 
Motin \textit{et al.}, among those, \citep {motin2019selection} utilised Empirical Mode Decomposition (EMD) and its variations to decompose PPG signals into their components. The authors then selected the Intrinsic Mode Functions (IMFs) devoid of artefacts and heart rate information, and applied the Principal Component Analysis (PCA) algorithm to the remaining components to extract respiratory rates. 

PPG signals can also be utilised to estimate vascular ageing. Dall'Olio \textit{et al.} \citep {dall2020prediction} proposed an algorithm for vascular ageing classification using PPG signals. The authors initially removed the trend in the PPG signal using a centered moving average. The processed PPG signal was then demodulated using a Hilbert transform, and the envelope was extracted. The final version of the PPG signal was obtained by dividing the demodulated signal by the envelope. A recursive peak detection method was subsequently applied to extract windows of 15 peaks, which were fed to a ResNet module for binary vascular ageing prediction. Additionally, a Support Vector Machine (SVM) model was constructed using features from metadata and features extracted from the   PPG signals. Hangsik \textit{et al.} \citep{shin2022photoplethysmogram} applied a five-layer convolutional neural network directly on the PPG pulses.

In 2009, Suzuki and Oguri conducted pioneering work on machine learning for blood pressure prediction from PPG \citep {suzuki2009cuffless}. The study utilised physiological features, including Pulse Width (PW), Transit Time (TW), Dicrotic Wave (DW), and Dicrotic Notch (Dn). An AdaBoost model was trained to predict systolic and diastolic blood pressures. Huang \textit{et al.} \citep{huang2022mlp} transferred the MLP-Mixer from the computer vision domain to the biomedical signal domain to predict blood pressure waveform using ECG and PPG signals with full MLP architecture.

Although the existing works have demonstrated good performance, several limitations prevent their more widespread use in practice. These include:
\begin{itemize}
    \item A considerable amount of preprocessing based on manually designed rules is required, such as a manual selection of intrinsic mode functions.\
    \item Existing algorithms are sensitive to parameter choice and may not generalise well across different datasets.\
    \item The majority of approaches employ the amplitude of the PPG signal as an input feature, which can be easily influenced by factors such as age, skin thickness, and skin tones.\
    \item Numerous methods directly work with features extracted from temporal domain PPG signals, making them highly sensitive to signal quality and effects of affine transformations.\
    \item Many existing solutions necessitate complex feature extraction operations.\
    \item Some algorithms seek to apply computer vision algorithms in the field of biomedical signal processing, which may result in unavoidable compromises, as these algorithms are primarily designed for processing 2D images.
\end{itemize}

To address these issues, we propose an algorithm that combines the amplitude-independent structure features and transfer learning techniques, named the VGTL-net. By virtue of the visibility graph approach, the time series of the PPG signal is transferred into a complex graph network which preserves the structural information while discarding the amplitude information. This makes it possible to employ the corresponding graph adjacency matrices (serving as images) in conjunction with transfer learning. And in this way, the VGTL-net can benefit from:
\begin{itemize}
    \item Use of only geometric information in the PPG signal, making the approach amplitude-independent;\ 
    \item Robustness to affine transformations by virtue of the visibility graph;\
    \item Minimal preprocessing requirements;\
    \item Good generalisation across different tasks and datasets;\
    \item Interpretation on how characteristics of the PPG signals are reflected in the visibility graph;\
    \item Minimal feature extraction and prerpocessing requirements;\
    \item Interpretation on how characteristics of the PPG signals are reflected in the visibility graph;\
     \item Information fusion through the RGB channels.
\end{itemize}

\begin{comment}
    Although such a conjoint treatment of graphs and transfer learning has been investigated before \citep{wang2021cuff}, the VGTL-net is entirely original in that it provides an in-depth analysis of the virtues of combining graphs and transfer learning, including:
\begin{itemize}
    \item Robustness to affine transformations by virtue of the visibility graph;\
    \item Use of only geometric information in the PPG signal, making the approach amplitude-independent;\
    \item Minimal preprocessing requirements;\
    \item Good generalisation across different tasks and datasets;\
    \item Interpretation on how characteristics of the PPG signals are reflected in the visibility graph;\
    \item Information fusion through the RGB channels.\
\end{itemize}

\end{comment}

This article is structured as follows. In Section \ref{sec:method}, we first give examples on how characteristics of the PPG signals are reflected in the visibility graph. Then, we introduce the proposed VGTL-net algorithm and outline the visibility graph and transfer learning techniques. In Section \ref{sec:exp}, we conduct experiments on two publicly available datasets using the proposed algorithm, demonstrating its generalisability in both tasks and datasets. Finally, in Section \ref{sec:conc}, we present conclusions discussing the performance and future directions.

\section{METHODOLOGY}
\label{sec:method}

\subsection{Signal To Graph: Visibility Graph}
\begin{figure}[h]
\centering
    \includegraphics[width=0.4\textwidth]{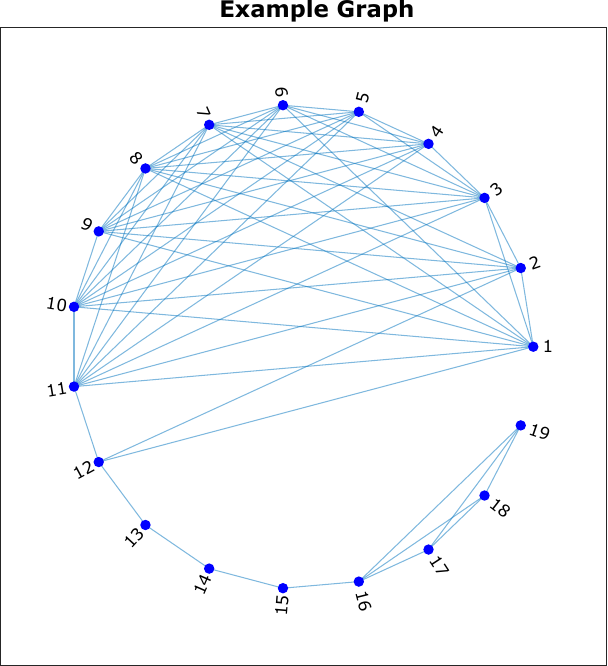}
    \caption{An example graph, with nodes which are indexed and form a circle. The lines connecting the nodes are the edges of the graph.}
    \label{fig:graph}
\end{figure}

\begin{figure}[h]
    \centering
    \begin{subfigure}[b]{0.3\textwidth}
        \includegraphics[width=\textwidth]{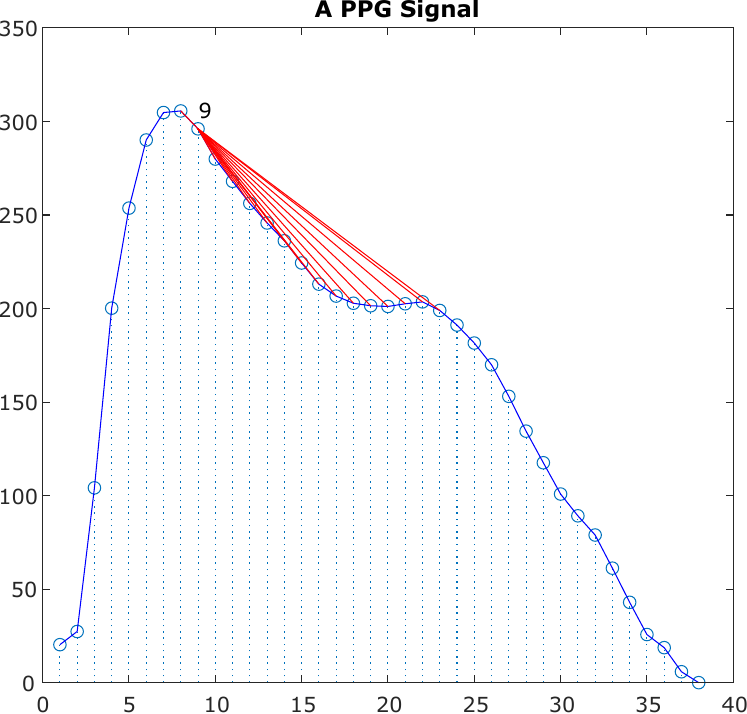}
        \caption{}
        \label{VG:subfig_a}
    \end{subfigure}

    \vspace{1em} % Optional space between subfigures

    \begin{subfigure}[b]{0.3\textwidth}
        \includegraphics[width=\textwidth]{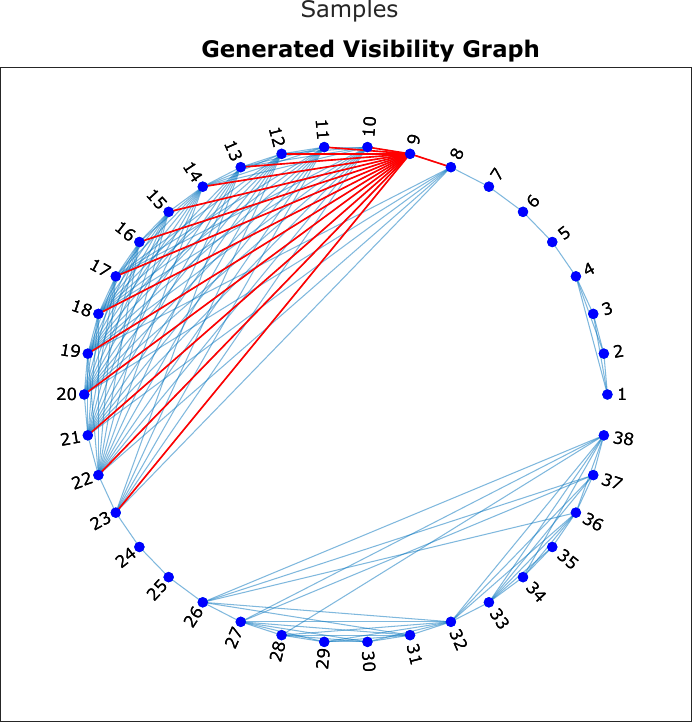}
        \caption{}
        \label{VG:subfig_b}
    \end{subfigure}
    \caption{An example of a visibility graph. (a): A PPG pulse and the red lines designate for the visibility of the signal sample 9. (b): The generated visibility graph and the red lines are the edges generated by the visibility of the signal sample 9. }
    \label{fig:VG}
\end{figure}

A graph $G = (V, E)$ is defined by a set of vertices $V$ connected by a set of edges $E$. Figure \ref{fig:graph} illustrates a graph where the vertices are represented as dots forming a circle, and the lines connecting them are the edges of the graph. For simplicity, all graphs in this work are undirected, which means that all edges are bi-directional. The visibility graph transforms a time series into a graph \citep{lacasa2008time}, whereby each signal sample $y_{i}$ is considered to be a vertex. There is an edge connecting two vertices if there exists a line connecting these two signal samples which does not intersect with another signal sample; this can be interpreted as one signal sample seeing another one without a third signal sample blocking the path. More formally, two vertices (signal samples), $y_{a}$ and $y_{b}$, are connected by an edge if
\begin{equation}
y_c<y_b+\left(y_a-y_b\right) \frac{t_b-t_c}{t_b-t_a}
\end{equation}
for any other signal sample $y_{c}$, where $t_{a}$, $t_{b}$ and $t_{c}$ are the time indexes corresponding to the signal samples $y_{a}$, $y_{b}$ and $y_{c}$. Since we consider undirected graphs only, all visibility is mutual, which means $y_{b}$ can see $y_{a}$ if $y_{a}$ can see $y_{b}$. Figure \ref{fig:VG} gives an example of the transformation from a sampled PPG cycle to a visibility graph. Figure \ref{VG:subfig_a} is the PPG signal and the red lines connect signal sample 9 with other signal samples that can be seen by it. These connections form edges that are red in the visibility graph, which is Figure \ref{VG:subfig_b}.

\begin{comment}    
\begin{figure}[h]
\centering
    \includegraphics[width=0.4\textwidth]{VG.pdf}
    \caption{An example of a visibility graph. plot upper part represent a PPG pulse and the lower plot is the generated visibility graph. The red lines in the upper plot designate for the visbility of signal sample 9 and the red lines in the lower plot are the edges generated by the visibility of signal sample 9. }
    \label{fig:VG}
\end{figure}
\end{comment}

\subsection{Graph To Image: Adjacency Matrix}
\begin{figure}[h]
    \centering
    \includegraphics[width=0.5\textwidth]{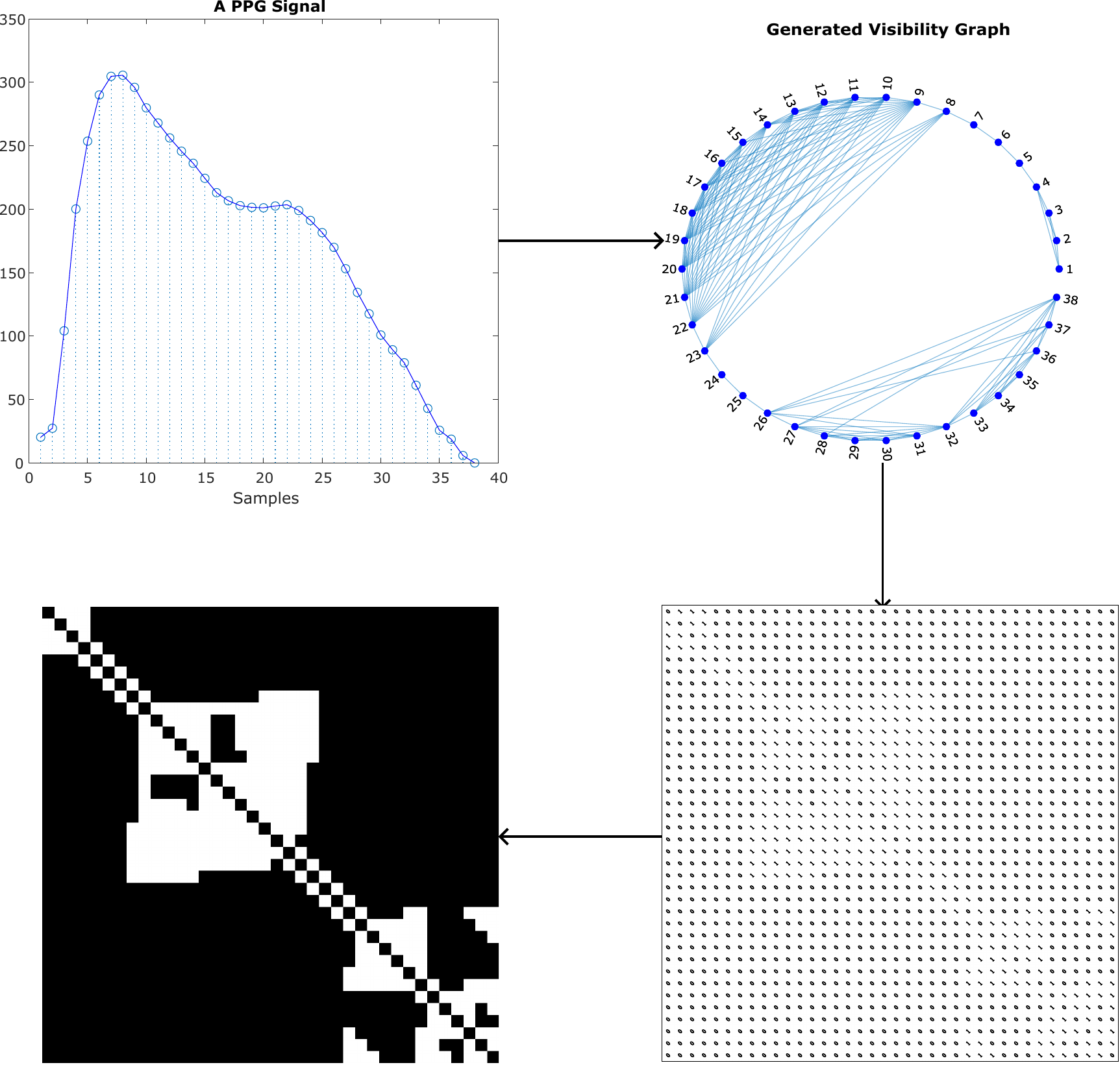}
    \caption{Pathway from a time series to an image. (top left): The original signal and the visibility between signal samples. (top right): The visibility graph generated from the input signal. (bottom right): The adjacency matrix which corresponds to the visibility graph. (bottom left): The black-and-white image generated from the adjacency matrix.}
    \label{fig:adj}
\end{figure}

The adjacency matrix is used to describe the connectivity of a graph. For a graph with $N$ vertices, an adjacency matrix $A$ has a size of $N \times N$. For a weighted graph, if two vertices $V_{i}$ and $V_{j}$ are connected, then $A_{i,j} = w_{i,j}$, while $A_{i,j} = 0$ if not connected. The weight $w_{i,j}$ may be calculated using some specific rules, for example, distance, correlation, etc. For an unweighted graph, $w_{i,j}$ will always be unity or zero. Note that the adjacency matrix can be considered as a grayscale image for a weighted graph and a black-and-white image for an unweighted graph, which is a form suitable as an input to a standard 2D convolutional neural network. Figure \ref{fig:adj} gives an example of a time series signal (top left), an unweighted graph generated from the signal (top right), and its corresponding adjacency matrix, viewed as both a matrix (bottom right) and an image (bottom left). Our proposed framework operates specifically with the adjacency matrix of the visibility graph. To this end, the time series PPG signals are first transformed into visibility graphs and then further transformed into images in the form of adjacency matrices.

\subsection{Visibility Graph: Invariance To Affine Transformation}
\begin{figure*}[h]
    \centering
    \begin{subfigure}[b]{0.2\textwidth}
        \includegraphics[width=\textwidth]{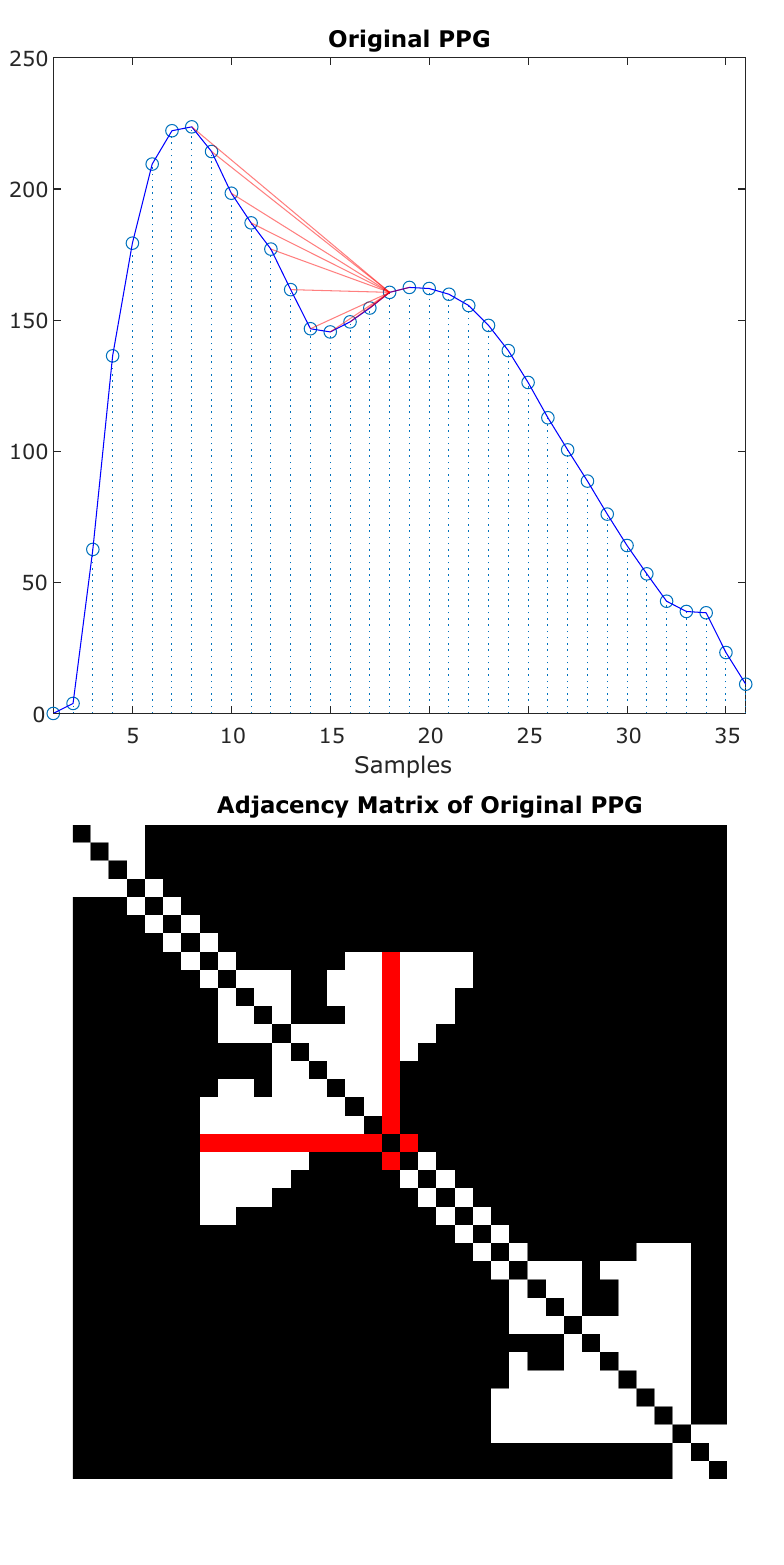}
        \caption{}
        \label{affine:subfig_a}
    \end{subfigure}
    \begin{subfigure}[b]{0.2\textwidth}
        \includegraphics[width=\textwidth]{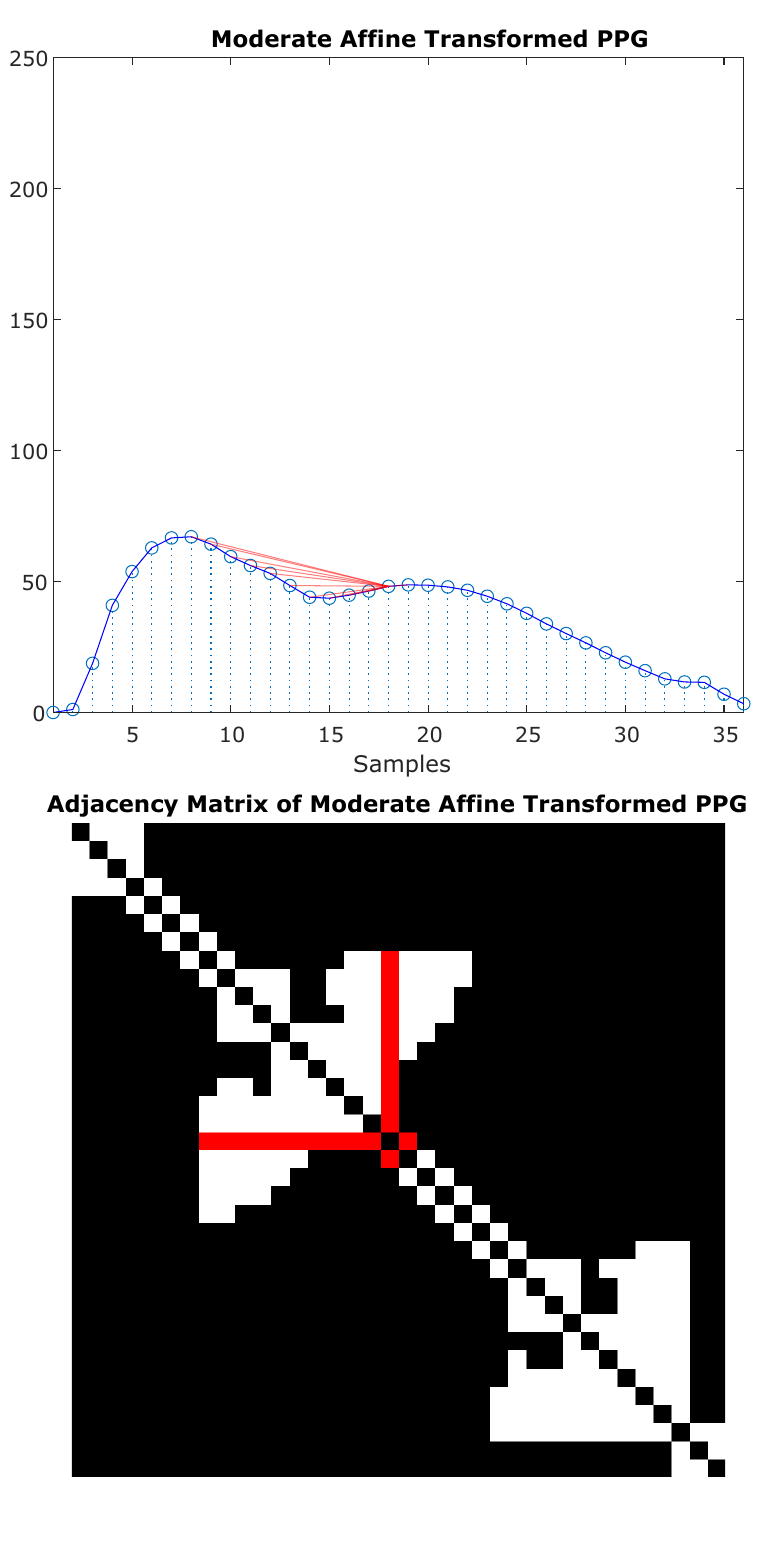}
        \caption{}
        \label{affine:subfig_b}
    \end{subfigure}
    \begin{subfigure}[b]{0.2\textwidth}
        \includegraphics[width=\textwidth]{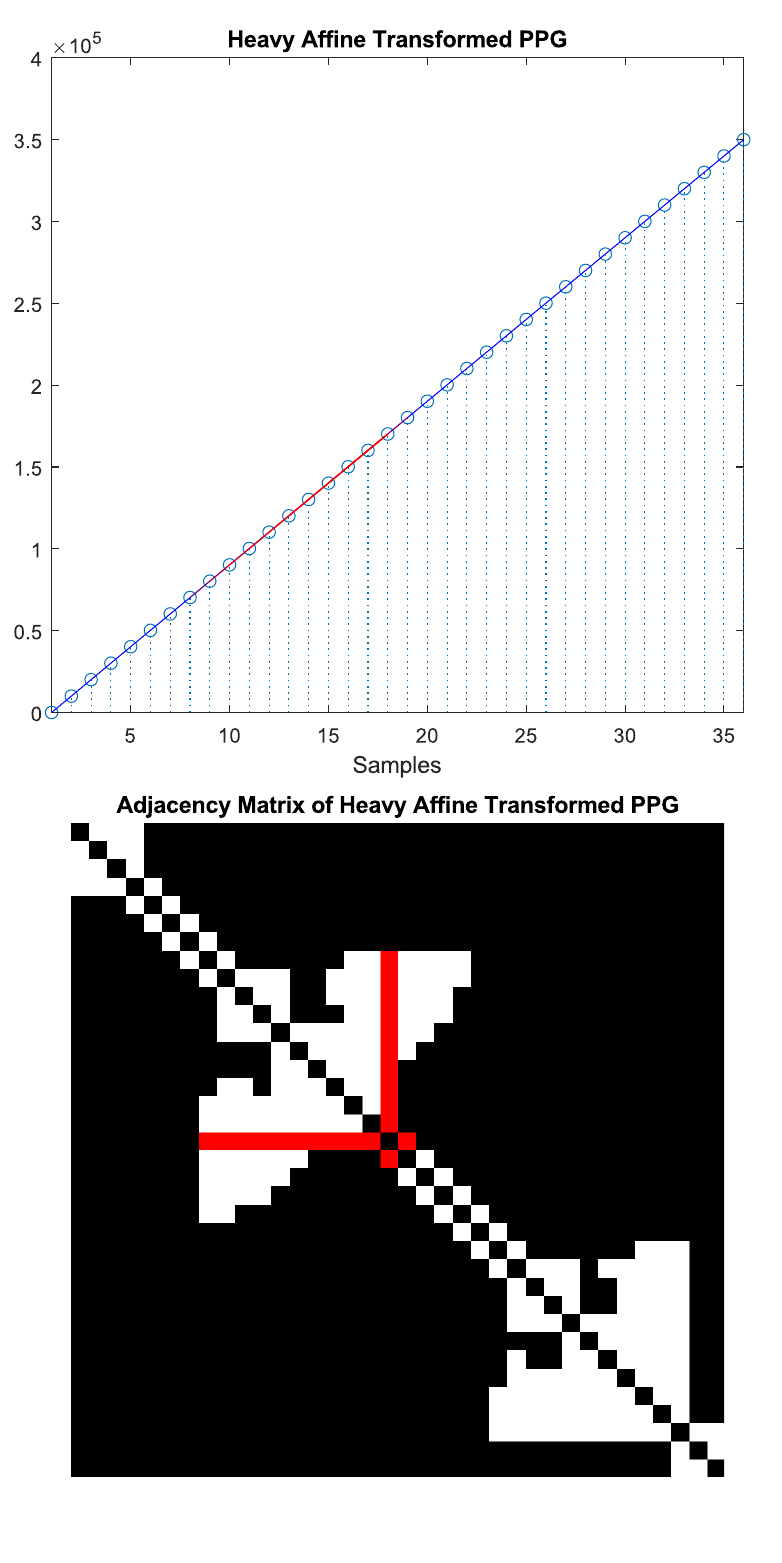}
        \caption{}
        \label{affine:subfig_c}
    \end{subfigure}
    \caption{(a): Adjacency matrices of the visibility graphs of the original PPG pulse. (b): The PPG pulse under moderate affine transformation. (c): The PPG pulse under heavy affine transformation. Although the PPG pulse becomes unidentifiable after affine transformations, the visibility graphs and the adjacency matrices remain the same, showing the visibility graph to be affine invariant. }
    \label{fig:affine}
\end{figure*}

The visibility graphs are invariant to affine transformations \citep{lacasa2008time}; in other words, the visibility graphs are invariant to horizontal and vertical scaling, horizontal and vertical translation, as well as the addition of any linear trend to the original signal. Figure \ref{fig:affine} provides an example, whereby the original signal (Figure \ref{affine:subfig_a}) is first moderately transformed (Figure \ref{affine:subfig_b}): 
\begin{equation}
    y(t) = 0.3x(t)
\end{equation}
then heavily transformed (Figure \ref{affine:subfig_c}): 
\begin{equation}
    y(t) = x(t) + 10000t + 13
\end{equation}
where $x(t)$ is the original PPG signal and $y(t)$ is the transformed signal.

Observe that even the heavily affine transformed signal has precisely the same visibility graph and the same adjacency matrix. For better illustration, we first found the signal sample that has the highest visibility in the original signal and denoted the index as $i$. Next, in all three signals, we connected the $i_{th}$ signal sample with the signal samples that can be seen by them. Then, we highlighted the points in the adjacency matrices generated by the connections in red. It is evident that with different levels of affine transformations, the visibility relationship does not change, and thus the visibility graph does not change either.

\begin{comment}
\begin{figure*}[htbp]
    \centering
    \includegraphics[width=0.9\textwidth, height=0.7\textheight, keepaspectratio]{PPG_affine.pdf}
    \caption{Adjacency matrices of the visibility graphs of the original PPG signals (in red) and the PPG signals under moderate and heavy affine transformation (in green). The adjacency matrices of the visibility graphs of these signals are identical, which demonstrates the affine-invariant of the visibility graph domain.}
    \label{fig:affine}
\end{figure*}
\end{comment}

\subsection{The PPG signal as a Visibility Graph}
We now demonstrate the effects of various breathing patterns, heart rates, and age on the adjacency matrices of the visibility graphs of the corresponding PPG signals. The examples show that visibility graphs can capture the structural characteristics of PPG signals, providing a foundation for further analysis. These results suggest that visibility graphs could potentially be used to analyse PPG signals and extract important physiological information.

\subsubsection{Breathing Modulation}
The PPG signal, a quantification of changes in blood volume through the absorbance of light, is subject to modulation by alterations in respiratory patterns. These modulations stem from variations in venous return, stroke volume, tissue volume, and respiratory sinus arrhythmia \citep{pimentel2015probabilistic, davies2022wearable}. The impact of respiratory activity modulation on PPG signals can be categorised into three distinct groups: amplitude modulation (AM), baseline wandering (BW), and frequency modulation (FM). Notably, BW modulation, which arises from fluctuations in venous pressure, is the most readily discernible characteristic within the raw PPG signal \citep{davies2022wearable}.

\begin{figure*}[h]
    \centering
    \begin{subfigure}[b]{0.8\textwidth}
        \includegraphics[width=\textwidth]{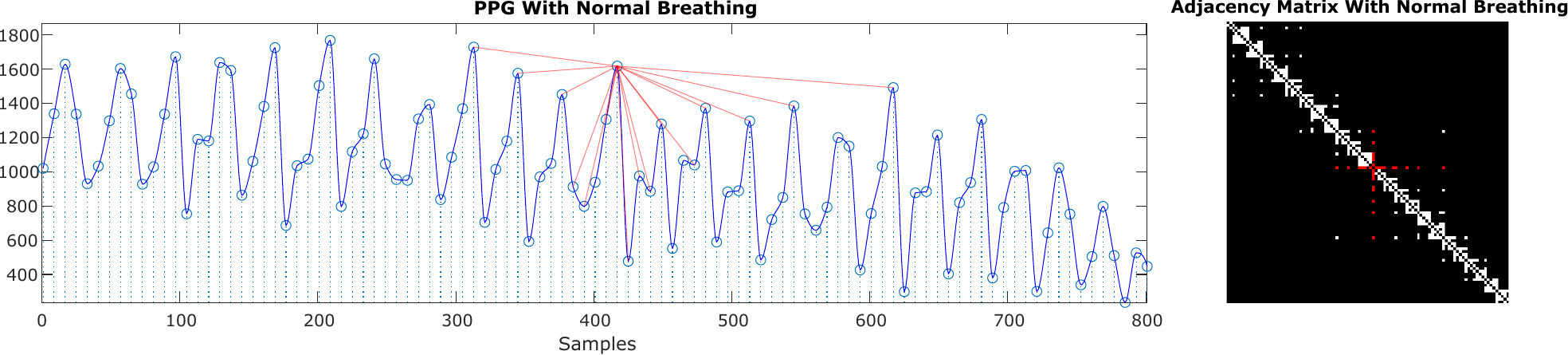}
        \caption{}
        \label{breathe:subfig_a}
    \end{subfigure}

    \vspace{1em} % Optional space between subfigures

    \begin{subfigure}[b]{0.8\textwidth}
        \includegraphics[width=\textwidth]{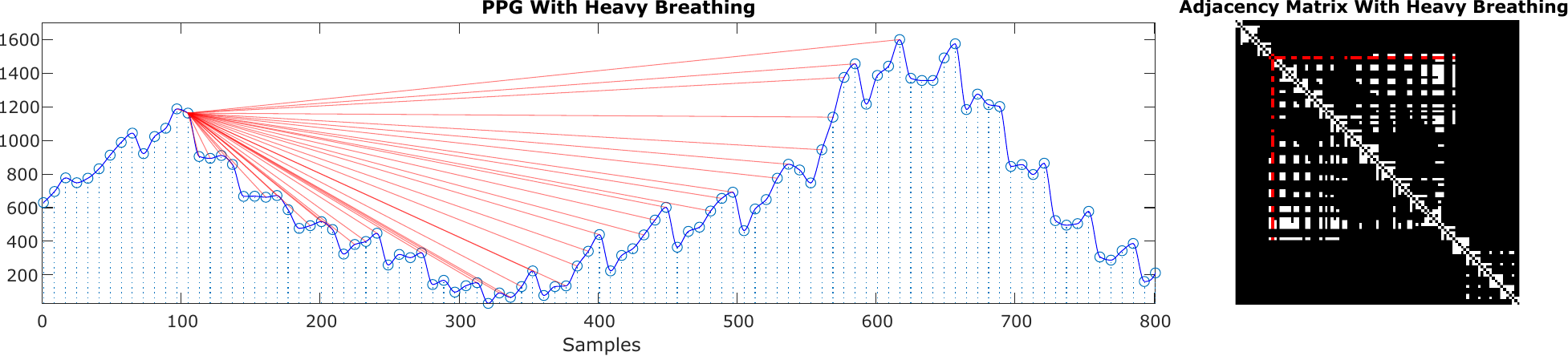}
        \caption{}
        \label{breathe:subfig_b}
    \end{subfigure}
    \caption{Example PPG waveforms and the corresponding adjacency matrices. (a): Normal breathing PPG signal and the adjacency matrix; (b): Heavy breathing PPG signal and the adjacency matrix. Red lines connect the signal sample with the highest visibility and the samples can be seen by it. The red pixels in the adjacency matrices represent edges introduced by the red lines. The baseline wandering introduced by heavy breathing gives some signal samples more visibility, thus referring to more red lines and broader wing-shape elements in the adjacency matrix. }
    \label{fig:breathe}
\end{figure*}

To investigate the effects of deep breathing on the PPG signal, we conducted an experiment whereby a subject initially breathed normally and then commenced deep breathing. Figure \ref{fig:breathe} depicts the resulting PPG signal segments and their corresponding adjacency matrices. Observe the BW modulation, a sinusoidal waveform shape, in the PPG signal. This modulation introduced additional peaks that were not present in the normal breathing PPG signal, thus giving certain pulses of the PPG signal additional visibility, which was manifested as wing shapes in the adjacency matrix. The signal samples with the highest visibility in both normal and heavy breathing were identified and were connected to the signal samples that can be seen by them in red lines. These connections were denoted in the adjacency matrices as red. It can be concluded that the heavy breathing introduced high visibility to some signal samples which can be seen as large wing shapes in the adjacency matrices.

\begin{comment}
\begin{figure}
    \centering
    \includegraphics[width = 0.5\textwidth]{Breathing.pdf}
    \caption{Example PPG waveforms and the corresponding adjacency matrices of normal breathing (left) and deep breathing (right).}
    \label{fig:breathe}
\end{figure}
\end{comment}

\subsubsection{Heart rates}
\begin{figure}[H]
    \centering
    \begin{subfigure}[b]{0.23\textwidth}
        \includegraphics[width=\textwidth]{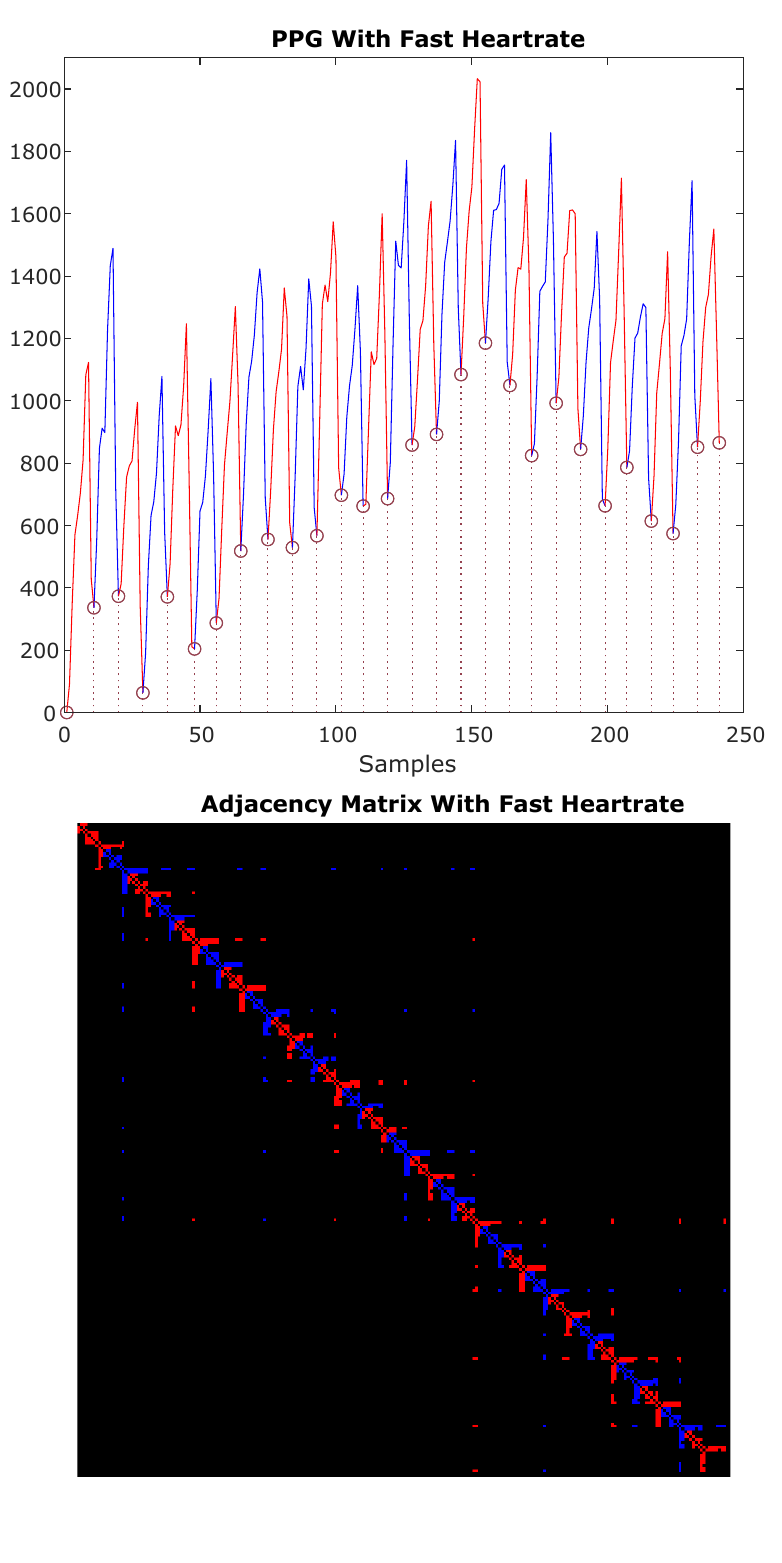}
        \caption{}
        \label{heart:subfig_a}
    \end{subfigure}
    \hfill
    \begin{subfigure}[b]{0.23\textwidth}
        \includegraphics[width=\textwidth]{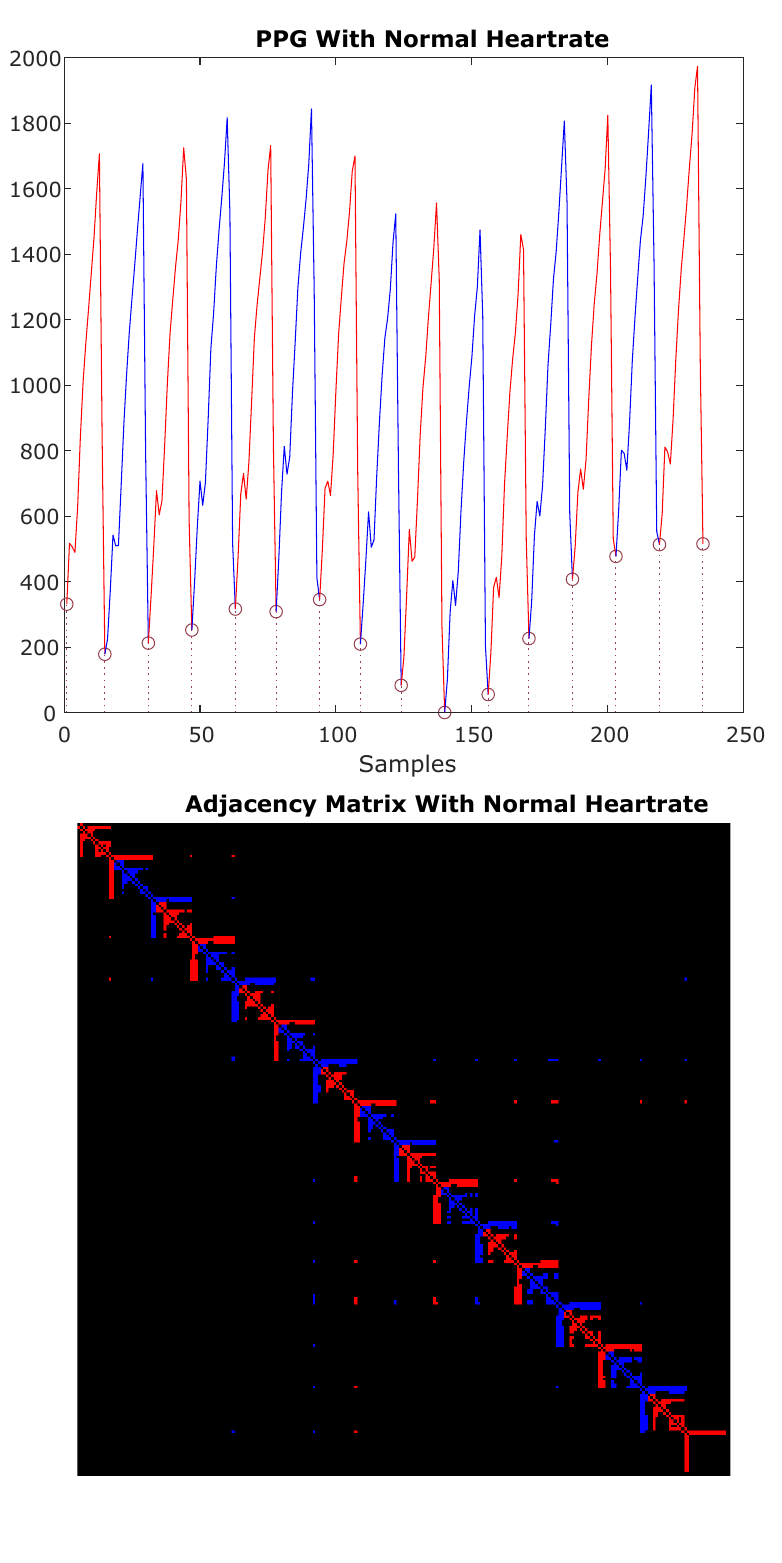}
        \caption{}
        \label{heart:subfig_b}
    \end{subfigure}
    \caption{Two segments of PPG signal with different heart rates and their corresponding adjacency matrices generated from the visibility graphs. (a): fast heart rate; (b) low heart rate. The wing shapes in the adjacency matrices share the same colour as the corresponding pulses for visualisations. It is clear that the number of wing shapes increases for a fast heart rate PPG signal with the same length as the normal rate PPG signal. }
    \label{fig:heart}
\end{figure}

An additional easily observable feature of PPG is the heart rate (HR). For illustration, the PPG signal of the subjects was initially recorded at rest and subsequently after exercising. Figure \ref{fig:heart} shows the differences in the corresponding adjacency matrices of the PPG segments with different heart rates. The wing-shaped elements along the diagonal signify peaks in the PPG signal. The individual pulses are represented alternately as red and blue. The sequence of colours in the plot is red, blue, red, blue, and so on. The wing shapes in the adjacency matrices share the same colour as the corresponding pulses for visualisations. Observe that the number of wing shapes increases for fast heart rate PPG signal with the same length as the normal rate PPG signal.

\begin{comment}
\begin{figure}
    \centering
    \includegraphics[width = 0.5\textwidth]{Heartrate.pdf}
    \caption{Two segments of PPG signal with different heart rates (top panel) and their corresponding adjacency matrices generated from the visibility graphs (bottom panel). }
    \label{fig:heart}
\end{figure}
\end{comment}

\subsubsection{Vascular Ageing}
\begin{figure}[h]
    \centering
    \begin{subfigure}[b]{0.23\textwidth}
        \includegraphics[width=\textwidth]{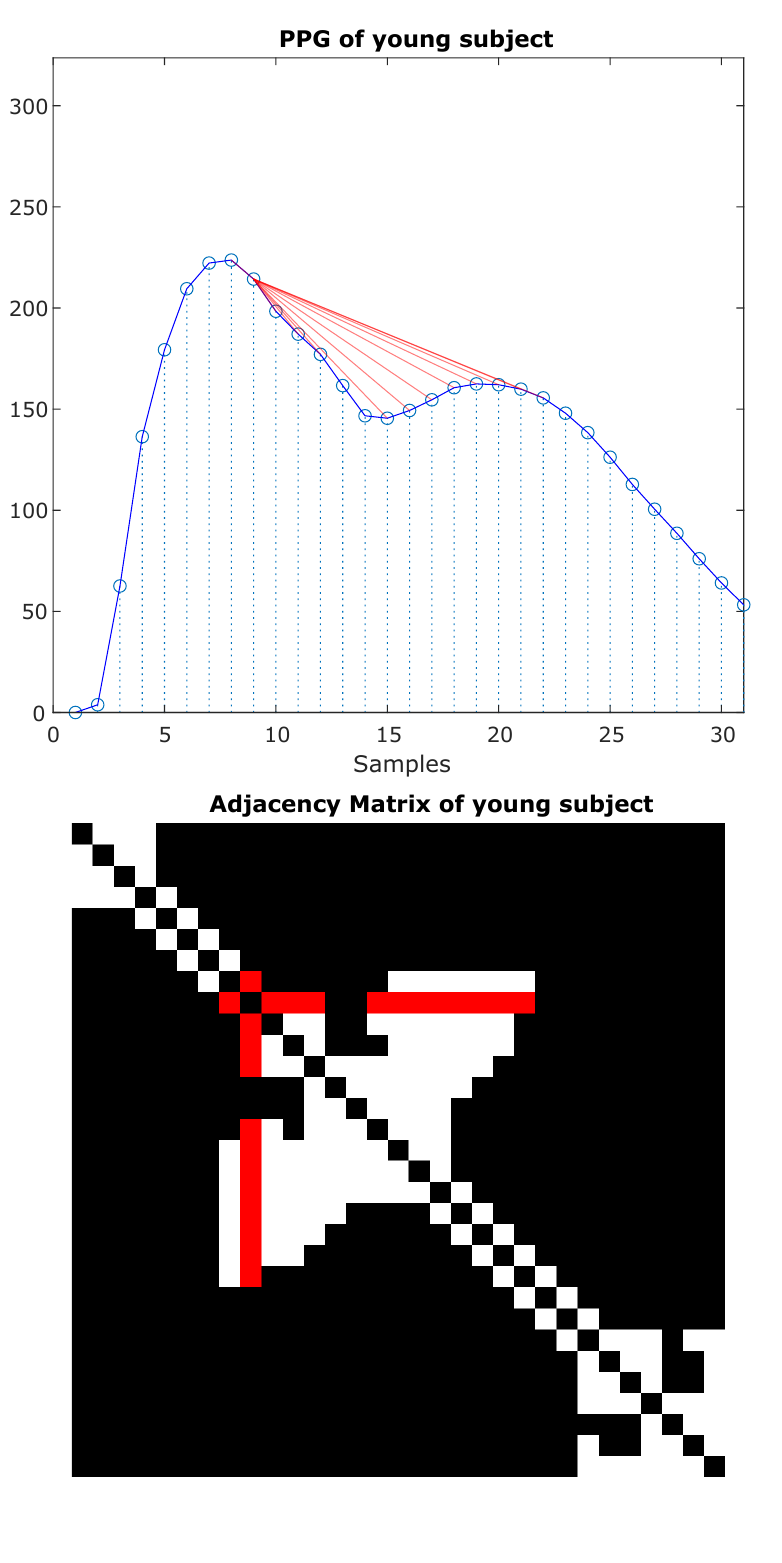}
        \caption{}
        \label{age:subfig_a}
    \end{subfigure}
    \hfill
    \begin{subfigure}[b]{0.23\textwidth}
        \includegraphics[width=\textwidth]{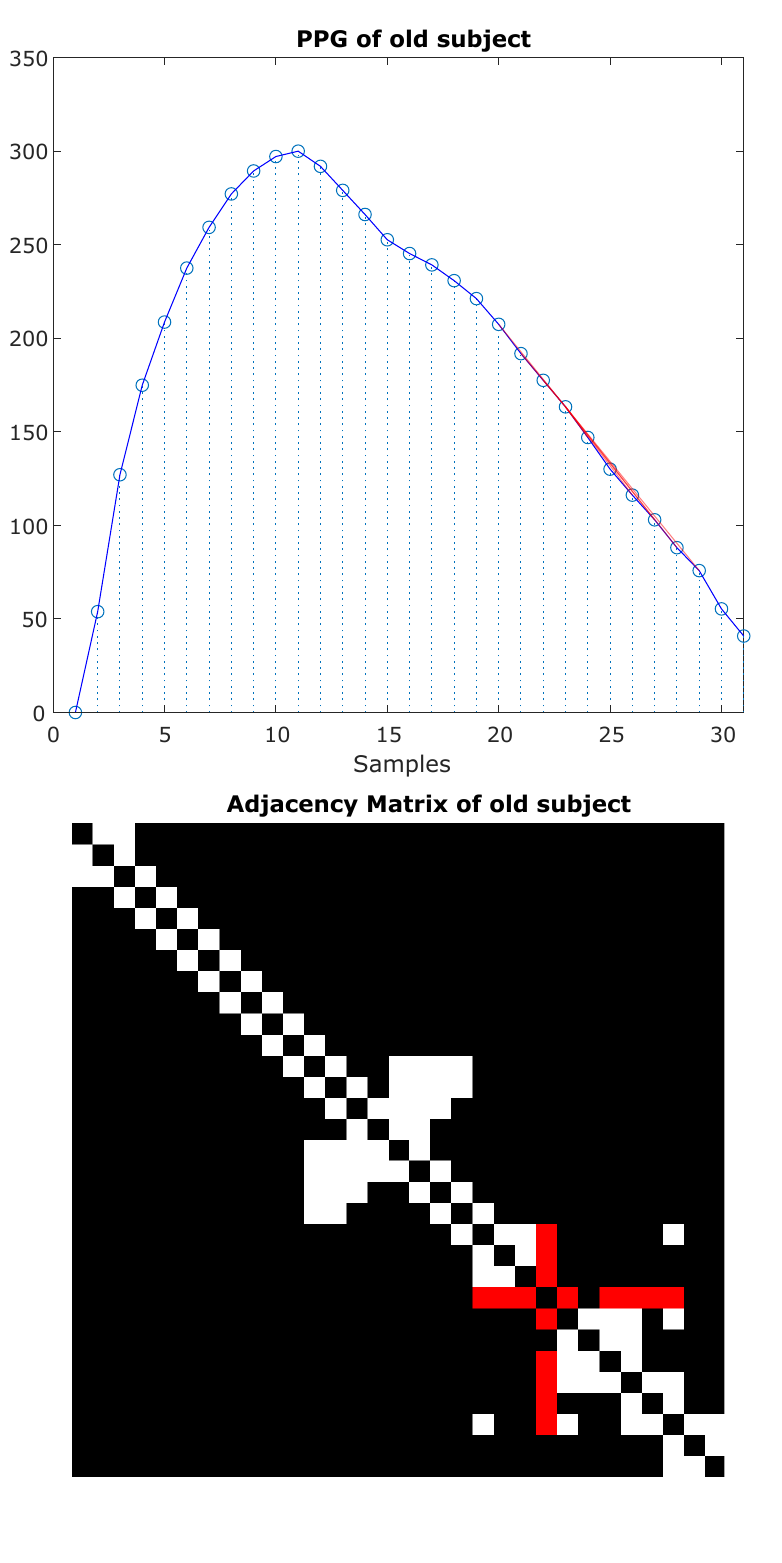}
        \caption{}
        \label{age:subfig_b}
    \end{subfigure}
    \caption{Age-related characteristics of PPG signals and the corresponding adjacency matrices. (a): 13 years old subject (b): 49 years old subject. It is obvious that the young PPG have larger wings due to the increased visibility compared with the old PPG}
    \label{fig:age}
\end{figure}

The shape of the PPG signal also varies with age, with the dicrotic notch less pronounced as age increases, yielding a smoother PPG signal \citep{yousef2012analysis}. Figure \ref{fig:age} offers an example of how the visibility graph can determine subjects' vascular ageing. Figure \ref{age:subfig_a} depicts the PPG signal and adjacency matrix of a subject aged 13, while Figure \ref{age:subfig_b} displays a 49 year old subject's PPG signal and adjacency matrix; note that the adjacency matrices exhibit significant differences. For a younger subject, the PPG signal is less smooth and has more notches, yielding the adjacency matrix with more white pixels which indicates that frequently two nodes can see each other. Conversely, due to the smoothness of the PPG signal of the older subject, most of the data points cannot see each other, producing an image less filled with white. To illustrate the increased visibility of a younger subject, the signal samples with the highest visibility in both young and old PPGs are selected and connected with the signal samples they could see using red lines. The red lines are also reflected in the adjacency matrices in red points. It is obvious that the PPG of a young subject has larger wings in the adjacency matrix due to the increased visibility compared with the PPG of the old subject.

\begin{comment}
\begin{figure}
    \centering
    \includegraphics[width = 0.5\textwidth]{age.pdf}
    \caption{Age related characteristics of PPG signals and the corresponding adjacency matrices. Left Panels: 13 years old subject Right Panels: 49 years old subject.}
    \label{fig:age}
\end{figure}
\end{comment}

In the previous section, we illustrated that to a certain extent physiological activities could be reflected in the adjacency matrices of the corresponding visibility graph. We next introduce the concept of transfer learning and show how it can be applied to visibility graphs.

\subsection{Transfer Learning}
Deep learning has achieved impressive results in the field of computer vision. However, state-of-the-art models often require large amounts of data for training, leading to the issue of data dependence. Transfer learning offers a solution to this problem by training models on large datasets to obtain a set of pre-trained weights, which are then fine-tuned on smaller datasets. This significantly reduces the required amount of new data and allows large deep learning models to be applied to a broader range of fields and scenarios \citep{tan2018survey}. Our study employed three pretrained models based on emperical tests, the sizes of the datasets and the nature of the tasks: the VGG19 model \citep{simonyan2014very} , the Coatnet \citep{dai2021coatnet}, and the Convnext v2 \citep{woo2023convnext} . All three pretrained models were trained on the ImageNet dataset \citep{deng2009imagenet}.
\begin{comment}
    The VGG19 model is a Convolutional Neural Network (CNN) with 19 layers, comprising 16 convolutional layers and three fully connected layers. It is equipped with five maxpooling layers and a final Softmax activation function \citep{simonyan2014very}. Depending on the task at hand, the output layer, which is the final fully connected layer, can be altered to meet the desired output dimension. The VGG19 model was initialised with weights trained on the ImageNet dataset \citep{deng2009imagenet}.    
\end{comment}

\subsection{Proposed Learning Scheme: VGTL-net}
The proposed learning scheme, termed the VGTL-net (Visibility Graph and Transfer Learning), is introduced with the aim of being capable of classification and regression tasks of arbitrary size. It comprises three steps: i) the PPG signal (or any other signal) is segmented into fixed-length windows or a fixed number of pulses; ii) the segmented PPG signals are then transformed into the corresponding adjacency matrices and further considered as grayscale or black-and-white images; iii) the adjacency matrices images are fed to the three channels of the pretrained model as an RGB image to give the final output.

\section{Experiments and Results}
\label{sec:exp}
We conducted two experiments on PPG signals, to validate the proposed VGTL-net approach. The first experiment considered the prediction of systolic and diastolic blood pressure, while the second experiment aimed to predict vascular ageing.

\subsection{Blood Pressure Estimation}
\subsubsection{Dataset}
\begin{figure}[h]
    \centering
    \includegraphics[width = 0.5\textwidth]{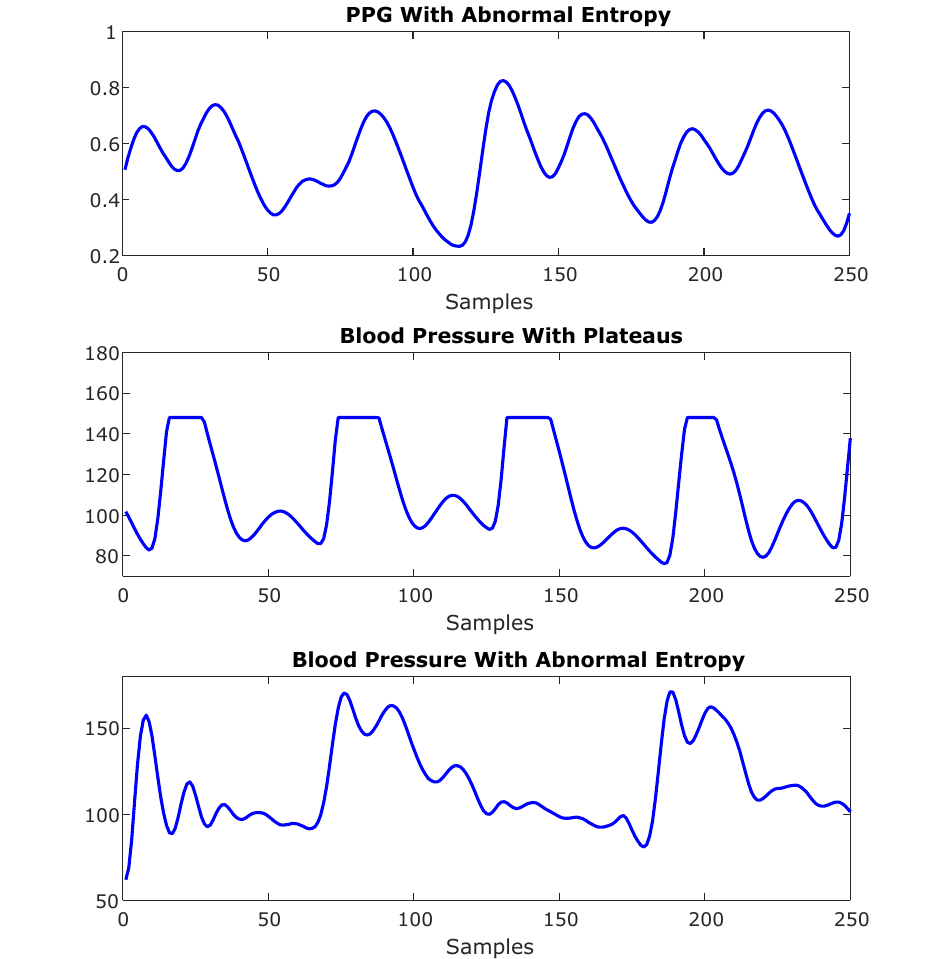}
    \caption{Top Row: One example of PPG signal with abnormal entropy; Middle Row: One example of blood pressure signal with plateaus; Bottom Row: One example of blood pressure signal with abnormal entropy.  }
    \label{fig:bad}
\end{figure}

The dataset came from the University of California Irvine (UCI) Machine Learning Repository which is a preprocessed version of MIMIC II dataset \citep{dataset}. The dataset contains PPG, ECG and arterial BP (ABP) signals all sampled at 125Hz. The dataset was further processed to select acceptable records:
\begin{itemize}
    \item Sample entropies were calculated for all trails and trials did not fulfil selected thresholds were removed.
    \item The trials with accepted entropy values were segmented into windows of 224 sample points which is suitable for the pretrained models.\\\\
    \item A plateau detection was performed, windows with plateaus longer than five points were deleted.\\
    \item The systolic and diastolic blood pressures were extracted from the windows left. Windows with extreme blood pressure values were deleted. Since the MIMIC dataset was recorded on patients in ICU under critical conditions, the accepted values of blood pressure were relaxed ($ 80 < SBP < 200$,  $40 < DBP < 120$).
\end{itemize}
Table \ref{table:bp_hr_ranges} shows the statistics of the processed dataset.
\begin{table}[H]
\centering
\begin{tabular}{lcccc}
\hline
\textbf{} & \textbf{Min} & \textbf{Max} & \textbf{STD} & \textbf{Mean} \\
 & \textbf{(mmHg)} & \textbf{(mmHg)} & \textbf{(mmHg)} & \textbf{(mmHg)} \\
\hline
DBP & 50 & 119.9 & 10.6 & 66 \\
MAP & 60.4 & 168.4 & 13.5 & 90.2 \\
SBP & 80 & 199.9 & 22 & 134.2 \\
HR & 46.3 & 234.3 & 42.6 & 117.2 \\
\hline
\end{tabular}
\caption{BP and Heart Rate Ranges in the Database}
\label{table:bp_hr_ranges}
\end{table}
Figure \ref{fig:bad} shows three examples of bad waveforms that were removed. The top and bottom row are windows of PPG signals with abnormal entropy values. The middle row illustrates a typical example of blood pressure waveform with plateaus. 

\subsubsection{Aim of the Experiment}
The goal of this experiment was to predict the systolic and diastolic blood pressures from the PPG and ECG signals. The PPG and ECG signals were transformed into adjacency matrices and the model predicted the systolic and diastolic blood pressures based on images transformed from these adjacency matrices. 
\subsubsection{RGB Fusion}
As mentioned above, the PPG signal and ECG signal were both segmented into windows of 224 sample points. We can utilize the fact that the input image has three channels, namely the red, green and blue channels. Different information can be transformed into adjacency matrices for different channels.

\begin{figure}[H]
    \centering
    \includegraphics[width = 0.5\textwidth]{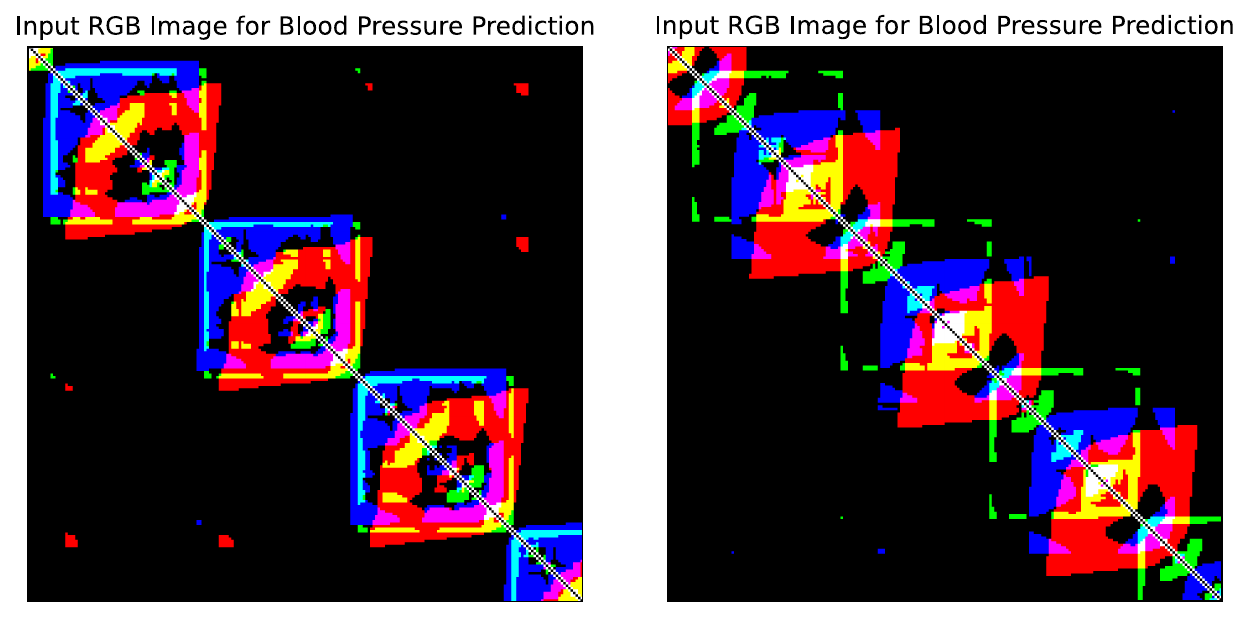}
    \caption{Two examples of input images for blood pressure prediction. It can be seen from the deviation between the red and green wings that the image on the right corresponds to PPG and ECG signals with higher pulse transit time compared with the signals the left image represents.}
    \label{fig:bp_rgb}
\end{figure}

The images were created with different information in the RGB channels. The pulse transit time calculated from ECG and PPG \citep{Kachuee2015CufflessHC} and the derivative of the PPG signal \citep{el2021cuffless} were often used as input features to estimate the blood pressures. To also encode this information in the generated images, the inputs to the red, green and blue channels are the PPG signal, the ECG signal and the first-order derivative of the PPG signal. The pulse transit time can be extracted by analysing the shift between the red and the green elements in the image, which are induced by the PPG and ECG signals respectively. In figure \ref{fig:bp_rgb}, it is manifest that the trial corresponding the right image clearly has a longer pulse transit time compared with the left image trial. The blue component corresponds to the first-order derivative of the PPG signal. 

As for the pretrained models, after empirical tests, two models outperform the others, the Convnext V2 \citep{woo2023convnext} and the Coatnet \citep{dai2021coatnet}.  We trained two VGTL-net models based on these two pretrained models. The classification layers of these two pretrained models were removed and the left parts were used as feature extractors. Then for each pretrained model, two MLPs of size (F×1024×512×1) were added as top layers to get the systolic and diastolic blood pressures, where F represent the feature size of that pretrained model. During training, the weights of the pretrained models and the top layers were updated. Furthermore, after the training of these two VGTL-nets, we trained a third VGTL-net model based on these two pretrained models. The trained Convnext V2 and the Coatnet feature extractors were extracted the the generated features were concatenated. Then two MLPs were added to form a VGTL-net. Finally, only the MLPs were trained to get further refined results.

\begin{comment}
    \subsubsection{Robustness-Enhancing Data Augmentation (REDA)}
We introduce the Robustness-Enhancing Data Augmentation (REDA) for the training stage. We have discovered that, by adding some bad quality trials that were removed from the prereprocessing stage, the model could converge faster and provides better performance. We did this by relaxing the plateau detection stage used for the curation of trainning set, which resulted in bad quality trials as in Figure \ref{fig:bad}. This forces the model to learn more robust features and reduce overfitting on the clean signals, resulting in a better overall performance.  
\end{comment}

\subsubsection{Results and Discussion}
\begin{figure}
    \centering
    \includegraphics[width=1\linewidth]{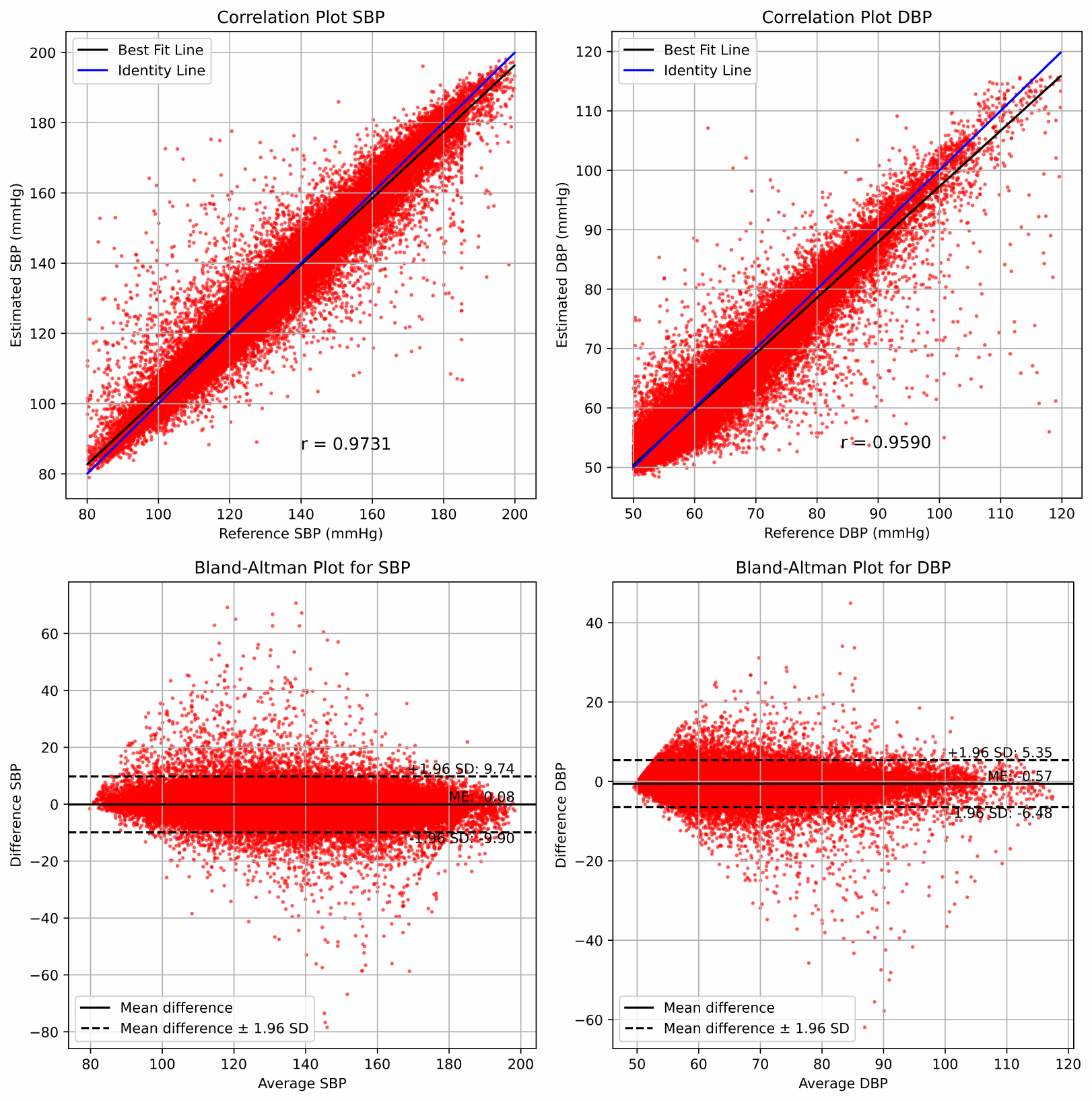}
    \caption{}
    \label{fig:all_results}
\end{figure}

Table \ref{table:ALL_results} shows the results of the systolic and diastolic blood pressure predictions in terms of mean absolute error (MAE), mean error (ME) and Pearson correlation coefficient. It contains other recent relevant research which also works on processed MIMIC II datasets and the three approaches we proposed, namely VGTL-net with Convnext v2, Coatnet and the concatenation of both of them as mentioned before. \begin{comment}\textbf{The AAMI standard requires the mean error and standard deviation to be within 5 mmHg and 8 mmHg which our method fulfils.}\end{comment} Table \ref{table:ALL_standards} gives the British Hypertension Society (BHS) blood pressure measurement grading system and our results along with other recent research. It is manifest that our method could achieve grade \textbf{A} for both systolic and diastolic blood pressures by significant margins. Also, our proposed methods perform overall better compared with all other research proposed in terms of MSE, MAE and absolute percentage difference. Figure \ref{fig:all_results} gives the correlation plots and the Bland-Altman plots for the SBP and DBP respectively.

\begin{figure}
    \centering
    \includegraphics[width=1\linewidth]{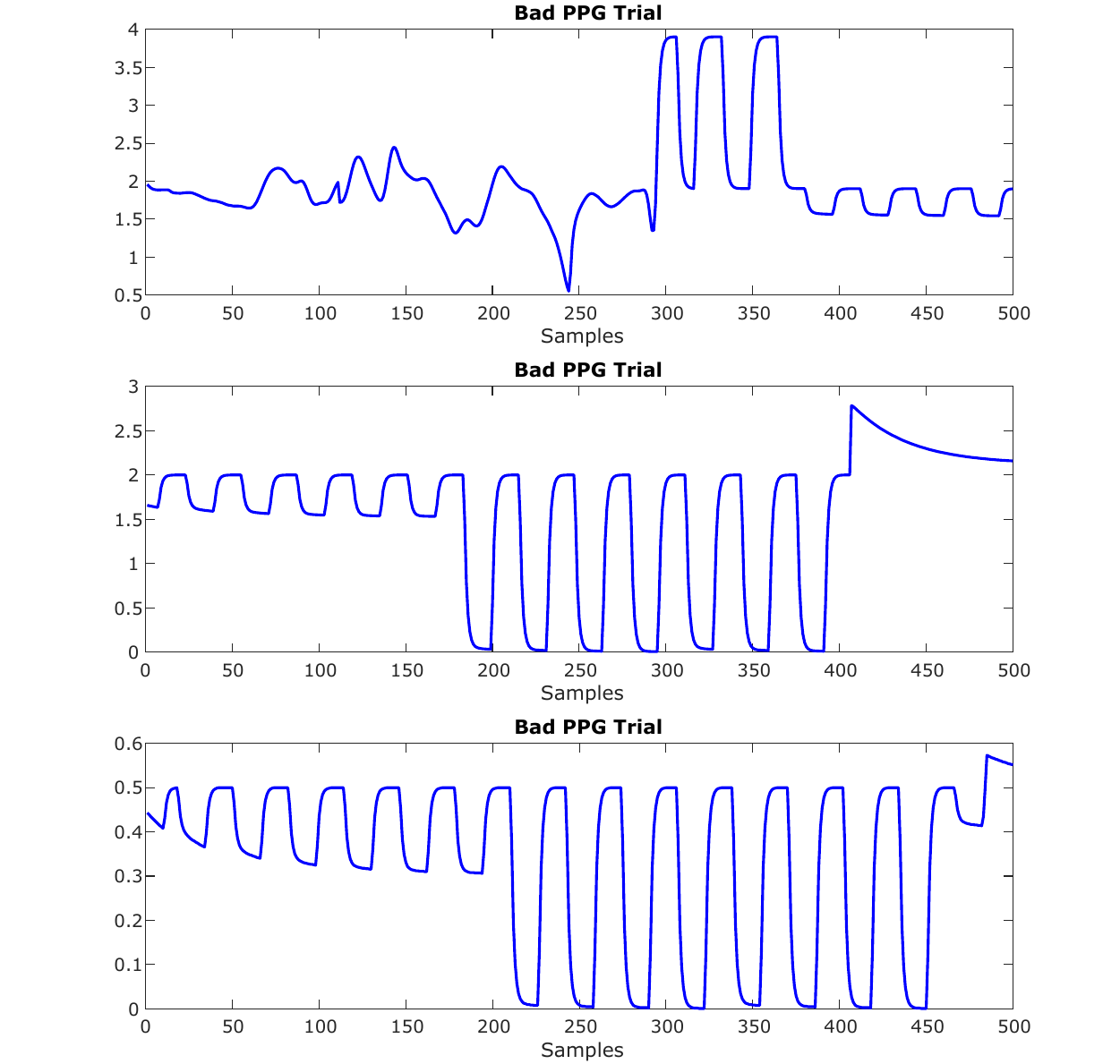}
    \caption{Three examples of heavily corrupted PPG signals.}
    \label{fig:BADPPG}
\end{figure}

It is important to mention that even after the deletion of corrupted trials, the quality of this dataset was severely undermined by the fact that the dataset was recorded on patients in critical conditions, with poor instrumentation and signal qualities, which affected the performance. This can be reflected by the abnormal statistic values in table \ref{table:bp_hr_ranges}. Figure \ref{fig:BADPPG} gives three examples of heavily corrupted PPG where the measurements that do not make sense at all. Furthermore, considering that the quality of the dataset is not optimal, the performance of our proposed algorithm could potentially be improved.  Moreover, no feature extracting was performed, which again indicated the low preprocessing requirement of VGTL-net.

\begin{comment}
    \begin{figure}
	\centering
	\includegraphics[width = 0.5\textwidth]{Predict_waveform.jpg}
	\caption{Two examples of predicting the blood pressure waveform using the proposed VGTL-net from PPG signal segments. The red lines denote the predicted blood pressure waveform while the black lines represent the ground truth.}
	\label{fig:Predict_waveform}
\end{figure}

\begin{figure}
	\centering
	\includegraphics[width = 0.5\textwidth]{bad_waveform.jpg}
	\caption{Two examples of predicting blood pressure waveform using the proposed VGTL-net from PPG signal segments when the ground truth blood pressure segments are of bad quality. The red lines denote the predicted blood pressure waveform and the black lines designate the ground truth. Observe that the VGTL-net was able to accurately predict blood pressure waveform even when the ground truth method was failing.}
	\label{fig:Bad_waveform}
\end{figure}
\end{comment}

\newcommand{\xmark}{\ding{53}}%
\newcommand{\customsize}{\fontsize{9}{13}\selectfont}
\begin{table*}
    \centering
    \customsize
    \begin{tabular}{cccccccc} \\
        \hline\\
        Model & \multicolumn{3}{c}{SBP} & & \multicolumn{3}{c}{DBP}\\
        \cline{2-4} \cline{6-8}
        & MAE (mmHg) & ME (mmHg) & \(\rho\) & & MAE (mmHg) & ME (mmHg) & \(\rho\) \\
        
        Rong \textit{et al.} \citep{rong2021multi} (PPG) & 5.59 & -1.13$\pm$ 7.25 & \xmark & & 3.36 & -0.14$\pm$ 4.48 & \xmark \\   

        El-Hajj and Kyriacou (PPG) \citep{el2021cuffless} & 4.51 $\pm$ 7.81 & -0.48 $\pm$ 9.16 & 0.89 & & 2.61 $\pm$ 4.41 & -0.49 $\pm$ 5.10 & 0.86\\   

        Slapnicar \textit{et al.} (PPG) \citep{slapnivcar2019blood} & 9.43 & \xmark & \xmark & & 6.88 & \xmark & \xmark \\ 

        Huang \textit{et al.} (PPG + ECG) \citep{huang2022mlp} &  3.52 $\pm$ 5.10 & -0.379 & 0.961 & & 2.13 $\pm$ 3.07 & -0.587 & 0.939 \\ 
        
        Miao \textit{et al.} (PPG + ECG) \citep{miao2020continuous} & \xmark & -0.11 $\pm$ 9.99  & 0.88 & & \xmark & -0.03 $\pm$ 6.36 & 0.71 \\ 
        
        Wang \textit{et al.} (PPG + ECG) \citep{wang2020end} &  3.95 $\pm$ 4.38 & \xmark & \xmark & & 2.14 $\pm$ 2.40 & \xmark & \xmark  \\ 

        Baker \textit{et al.} (PPG + ECG) \citep{baker2021hybrid} &  4.41 $\pm$ 6.11 & \xmark & 0.8 & & 2.91 $\pm$ 4.23 & \xmark & 0.85\\ 

        Esmaelpoor \textit{et al.} (PPG + ECG) \citep{esmaelpoor2020multistage} &  3.97 $\pm$ 5.55 & 1.91 & 0.954 & & 2.10 $\pm$ 2.84 & 0.67 & 0.950  \\ 

        Long \textit{et al.} (PPG + ECG) \citep{long2023bpnet} &  3.98 $\pm$ 4.62 & -0.17 & 0.9564 & & 2.33 $\pm$ 2.95 & -0.24 & 0.9340  \\ 
        
        Proposed Model (CoAtNet) (PPG + ECG) & 3.46 $\pm$ 4.22 & -0.19 $\pm$ 5.46 & 0.967  &  & 2.11 $\pm$ 2.51 & -0.02 $\pm$ 3.29 & 0.950 \\ 
        
        Proposed Model (ConvNeXt V2) (PPG + ECG) & 3.29 $\pm$ 4.20 & 0.03 $\pm$ 5.34 & 0.969 &  & 1.99 $\pm$ 2.49 & -0.03 $\pm$ 3.19 & 0.953 \\
        
        Proposed Model (Concatenate) (PPG + ECG) & 3.11 $\pm$ 3.92 & -0.07 $\pm$ 5.01 & 0.973 &  & 1.94 $\pm$ 2.37 & -0.5 $\pm$ 3.01 & 0.959 \\ 
        \hline\\
        
    \end{tabular}
    \caption{The numeric results of the VGTL-net and other works.}\label{table:ALL_results}
\end{table*}

\begin{table*}
    \centering
    \customsize
    \begin{tabular}{cccccccc} 
        \hline\\
        Model & \multicolumn{3}{c}{SBP} & & \multicolumn{3}{c}{DBP}\\
        \cline{2-4} \cline{6-8}
       & \(\leq 5 ~ mmHg\) & \(\leq 10 ~ mmHg\) & \(\leq 15 ~ mmHg\) & & \(\leq 5 ~ mmHg\) & \(\leq 10 ~ mmHg\) & \(\leq 15 ~ mmHg\) \\
        
        Rong \textit{et al.} \citep{rong2021multi} (PPG) & 54.1\% & 86.6\% & 94.9\% & & 82.9\% & 94.9\% & 98.3\% \\   

        El-Hajj and Kyriacou (PPG) \citep{el2021cuffless} & \xmark & \xmark & \xmark & & \xmark & \xmark & \xmark \\   

        Slapnicar \textit{et al.} (PPG) \citep{slapnivcar2019blood} & \xmark & \xmark & \xmark & & \xmark & \xmark & \xmark \\ 

        Huang \textit{et al.} (PPG + ECG) \citep{huang2022mlp} & 71.23\% &  92.28\% & 98.92\% & & 88.07\% & 98.65\% & 99.78\% \\ 

        Miao \textit{et al.} (PPG + ECG) \citep{miao2020continuous} & 50.07\% &  76.40\% & 90.39\% & & 65.66\% & 89.77\% & 96.63\% \\ 

        Wang \textit{et al.} (PPG + ECG) \citep{wang2020end} & \xmark &  \xmark & \xmark & & \xmark & \xmark & \xmark \\ 

        Baker \textit{et al.} (PPG + ECG) \citep{baker2021hybrid} & 67.66\% &  89.82\% & 96.82\% & & 82.79\% & 96.12\% & 99.58\% \\ 

        Esmaelpoor \textit{et al.} (PPG + ECG) \citep{esmaelpoor2020multistage} & 73.7\%  & 93.7\% & 97.7\% & & 92.9\% & 99.2\% & 99.9\% \\ 

        Long \textit{et al.} (PPG + ECG) \citep{long2023bpnet} & 74.47\%  & 95.52\% & 97.05\% & & 89.57\% & 97.70\% & 99.13\% \\ 
        
        Proposed Model (CoAtNet) (PPG + ECG) & 79.64\% & 94.71\% & 97.91\% & & 91.82\% & 98.39\% & 99.49\% \\ 
        
        Proposed Model (ConvNeXt V2) (PPG + ECG) & 81.63\% & 95.23\% & 98.00\% &  & 93.05\% & 98.53\% & 99.47\% \\
        
        Proposed Model (Concatenate) (PPG + ECG) & 83.07\% & 95.84\% & 98.28\% & & 93.52\% & 98.74\% & 99.54\% \\  

        \hline \\ 

        Grade & \multicolumn{6}{c}{ Absolute Difference between the Standard and Test results (\%) }\\
        \cline{2-7}\\
        A & & \multicolumn{4}{c}{80  \hfill 90  \hfill 95 }\\
        B & & \multicolumn{4}{c}{65  \hfill 85  \hfill 95 }\\
        C & & \multicolumn{4}{c}{45  \hfill 75  \hfill 90 }\\
        D & & \multicolumn{4}{c}{Worth than C}\\
        \hline\\

    \end{tabular}
    \caption{The British Hypertension Society standard (BHS) and the accuracy of the proposed VGTL-net and other works on blood pressure prediction.}\label{table:ALL_standards}
\end{table*}

\subsection{Vascular Ageing}
\subsubsection{Dataset}
The dataset used for vascular ageing prediction was the ''Real-World PPG Dataset'' from Mendeley \citep{Siam2019-qb}. This dataset consists of PPG signals for 35 healthy subjects, with 50-60 PPG signals per subject. The duration of each PPG signal is 6 seconds, with a sampling rate of 50 Hz. The ages of subjects range from 10 to 75 years old. The dataset is divided into the training, validation, and test datasets.

\subsubsection{Aim of The Experiment}
This experiment utilised the proposed VGTL-net scheme to predict the age of subjects based on their PPG signals. The prediction was conducted in the regression and classification settings. For the regression task, the goal of the VGTL-net was to predict the exact age of the subject, while for the classification task, the goal of the VGTL-net was to determine the age group of the subject, namely 0-20, 20-30, 30-40, and 40 +.

\subsubsection{Data Preparation and Model Architecture}
\begin{figure}[H]
    \centering
    \includegraphics[width = 0.35\textwidth]{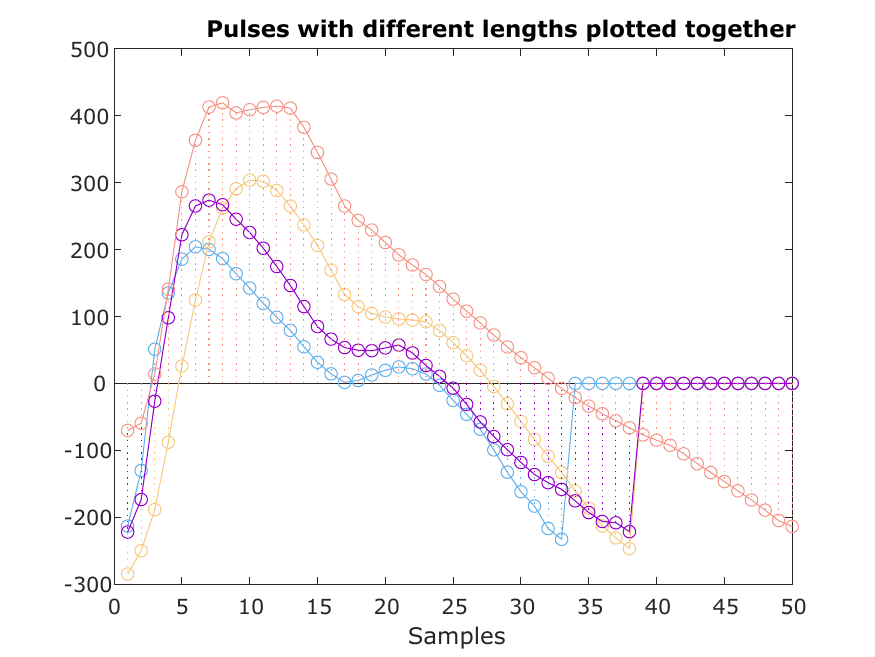}
    \caption{PPG pulses of different lengths visualised in the same graph. Pulses longer than 50 data points (orange) are truncated and pulses shorter than 50 data points (blue, purple and amber) are zero-padded.}
    \label{fig:pad_n_cut}
\end{figure}

The PPG signals were segmented into individual peaks, with every training, validation or test sample containing only one peak. As the heart rate can change both between and within subjects, an average length of 50 samples was selected (1 second). Segments longer than 50 samples were truncated, and those shorter than 50 were padded with zeros. Figure \ref{fig:pad_n_cut} illustrates such padding and truncation.

\begin{figure}[h]
    \centering
    \includegraphics[width = 0.25\textwidth]{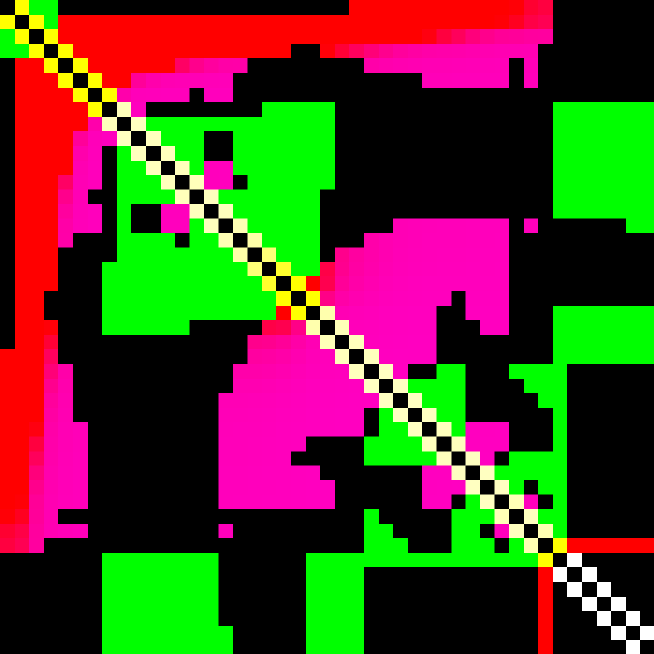}
    \caption{The RGB version of the input visibility graphs adjacency matrices. The RGB channels from Figure \ref{fig:Three} are combined to form this image. Different modes of information are fused and form an RGB image.}
    \label{fig:rgb}
\end{figure}

\begin{figure}[H]
    \centering
    \begin{subfigure}[b]{0.4\textwidth}
        \includegraphics[width=\textwidth]{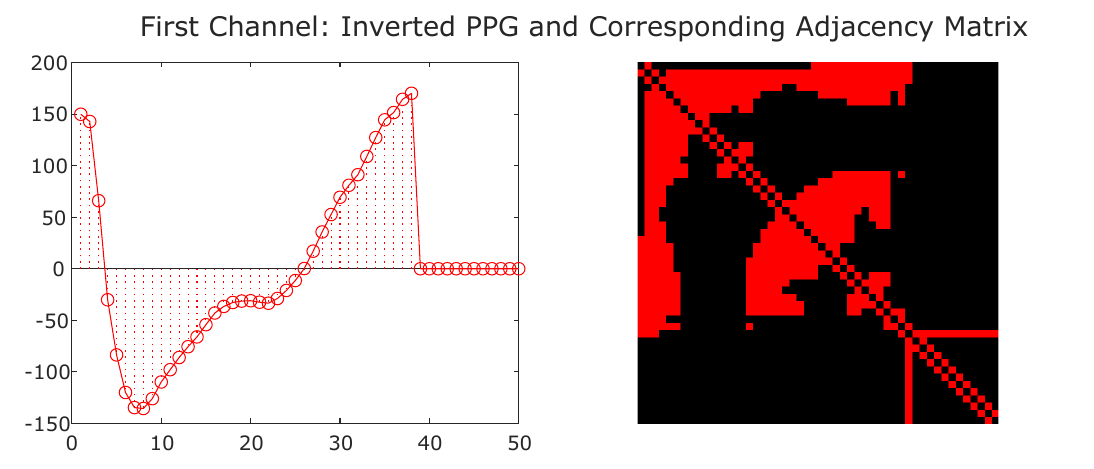}
        \caption{}
        \label{Three:subfig_a}
    \end{subfigure}

    \vspace{1em} % Optional space between subfigures

    \begin{subfigure}[b]{0.4\textwidth}
        \includegraphics[width=\textwidth]{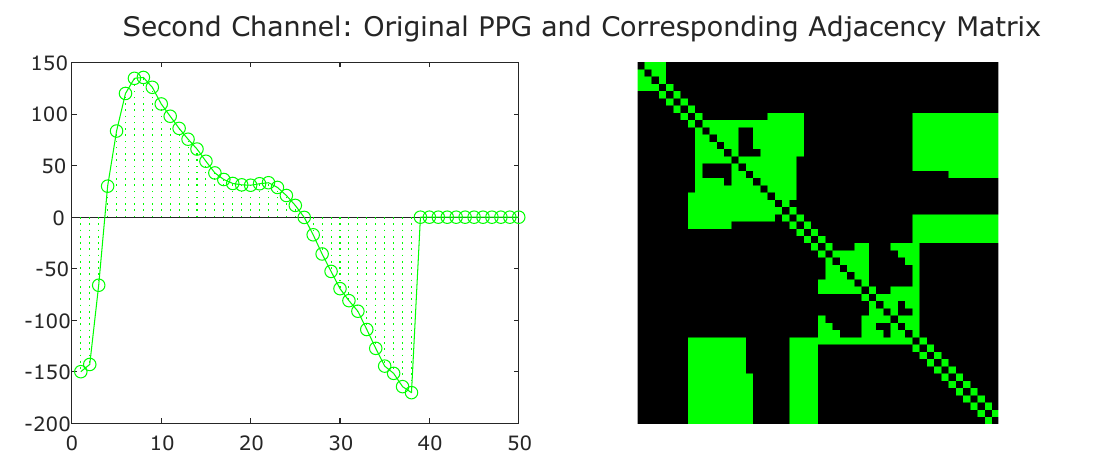}
        \caption{}
        \label{Three:subfig_b}
    \end{subfigure}

    \vspace{1em} % Optional space between subfigures

    \begin{subfigure}[b]{0.4\textwidth}
        \includegraphics[width=\textwidth]{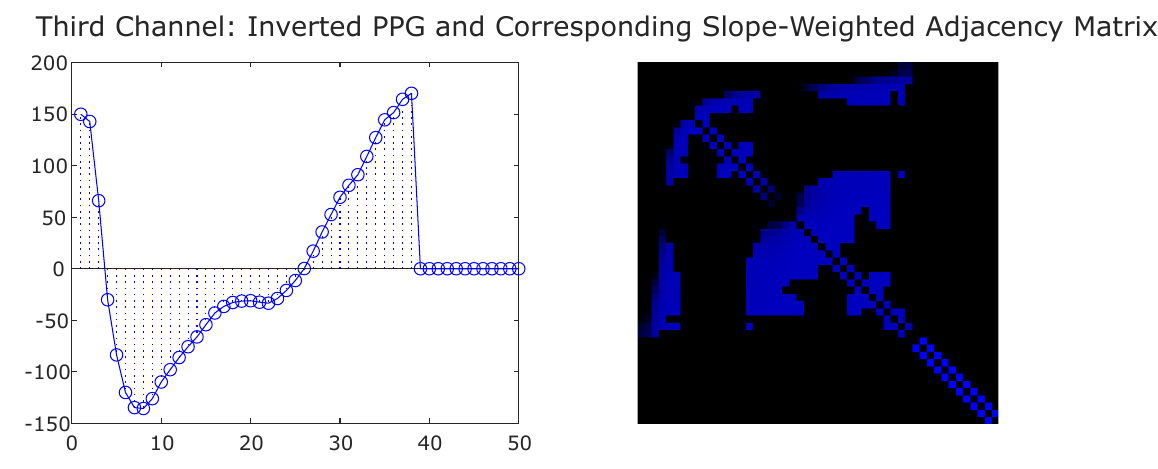}
        \caption{}
        \label{Three:subfig_b}
    \end{subfigure}
    
    \caption{Three channels of a PPG pulse and their corresponding adjacency matrices. (a): Red channel: the adjacency matrix generated from the PPG pulse. (b): Green channel: the adjacency matrix generated from the amplitude-inverted PPG pulse. (c): Blue channel: the adjacency matrix of a slope-weighted visibility graph generated from the amplitude-inverted PPG pulse.}
    \label{fig:Three}
\end{figure}

Next, the PPG pulses were transformed into visibility graphs as in Figure \ref{fig:graph}, and the corresponding adjacency matrices were converted into images to feed the VGG19 model which was pre-trained on the ImageNet dataset. Figure \ref{fig:Three} illustrates the three-channel image input which is required for the VGG19 model; the three inputs were the adjacency matrix generated from the PPG pulses, the adjacency matrix generated from the amplitude-inverted PPG pulses, and the adjacency matrix of a slope-weighted visibility graph generated from the amplitude-inverted PPG signal. For slope-weighted visibility graphs, the edge weights (the non-zero values of the adjacency matrices) depend on the slope of the line connecting the two data points. These different modes of information were fused by considering them as the RGB channels. With these channels as the RGB channels of a colour image, Figure \ref{fig:rgb} shows an example of a resulting input image of the VGG19 model, with no preprocessing required. The \textbf{findpeaks} function in \textbf{MATLAB} was used to extract pulses from the PPG signals. Then, the adjacency matrices were shuffled and divided into the training, validation, and test sets with sizes of 9,148, 3,050, and 3,049. Ten consecutive tests were performed using different training/validation/test distributions (different random seeds). The results were analysed by calculating the mean and standard deviation.

As for the model architecture, the VGG19 model was pre-trained on the ImageNet dataset with 1,000 classes. Thus, a fully connected top layer with the size of $1000 \times 4$, which gives a four dimension output suitable for classifying subjects into four age groups, was added to the VGG19 model for the classification task. In addition, the top layers of size $1000 \times 1$ and $1000 \times 2$ were included for the regression and binary classification task according to the required output dimension.

\begin{comment}
\begin{figure}[h]
    \centering
    \includegraphics[width = 0.45\textwidth]{Three.pdf}
    \caption{Three channels of a PPG pulse (Left) and their corresponding adjacency matrices (Right).}
    \label{fig:Three}
\end{figure}
\end{comment}

\subsubsection{Results and Discussions}
\begin{figure}[h]
    \centering
    \includegraphics[width = 0.4\textwidth]{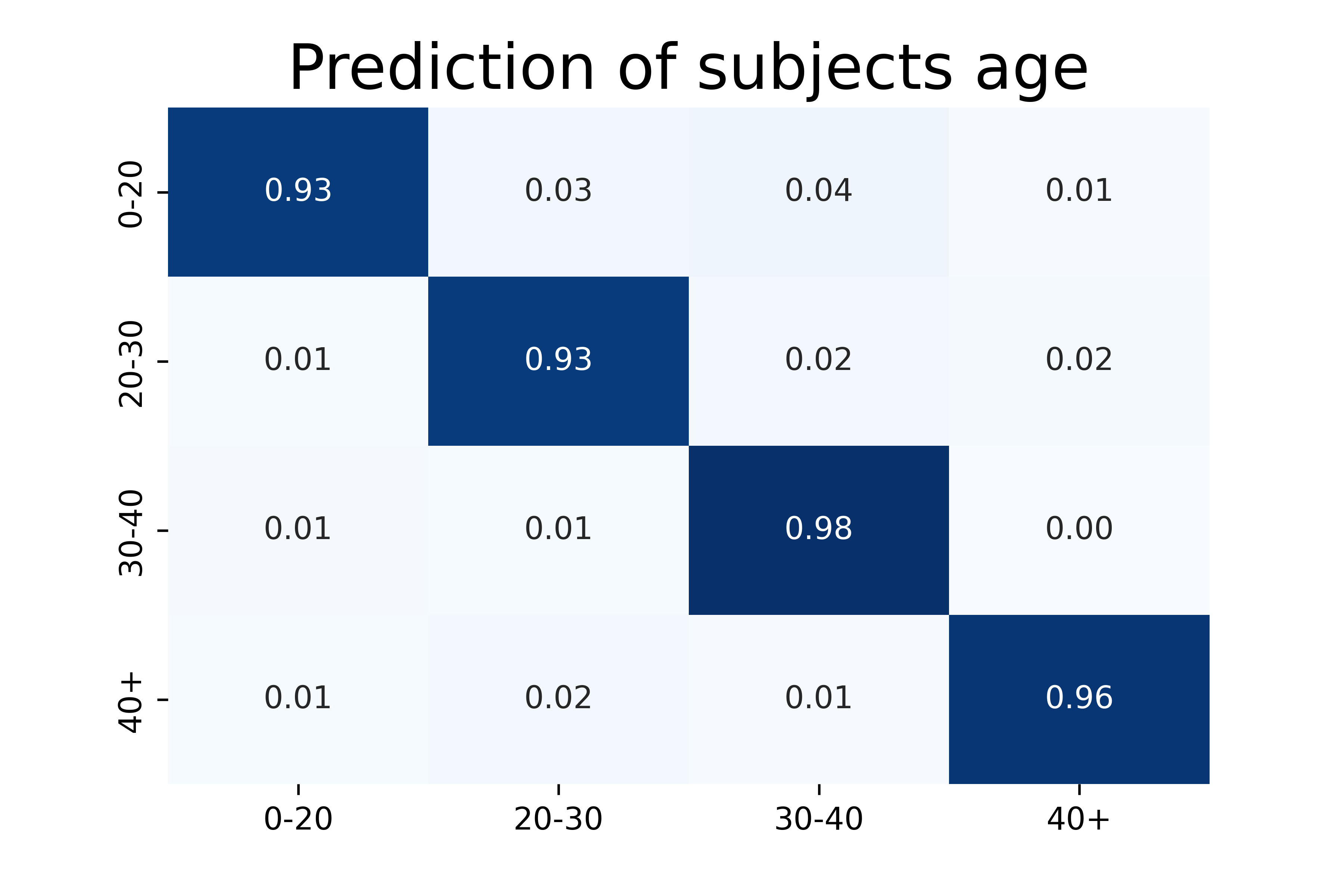}
    \caption{The confusion matrix of the performance of the VGTL-net on the four-class task of classifying subjects into age groups of 0-20, 20-30, 30-40, and 40+ years old}
    \label{fig:CM4}
\end{figure}

\begin{figure}[h]
    \centering
    \includegraphics[width = 0.4\textwidth]{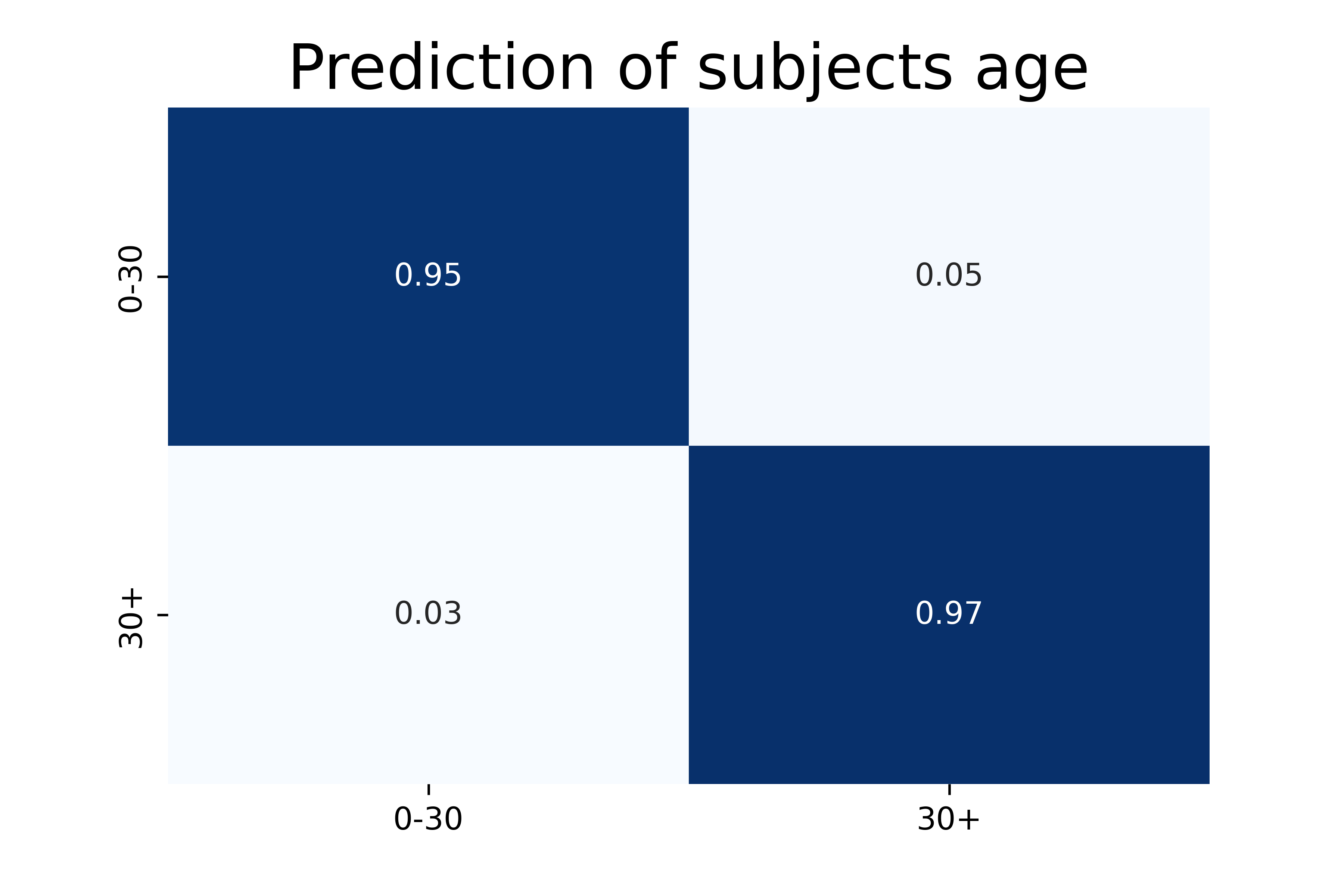}
    \caption{The confusion matrix of the performance of the VGTL-net on the binary task of classifying subjects into age groups of 0-30 and 30+ years old.}
    \label{fig:CM2}
\end{figure}

We compared the proposed methods for the classification task with that introduced by Dall'Olio \textit{et al.} \citep{dall2020prediction}, which performed a binary classification into the young and old. To make a fair comparison, we adopted the method in \citep{dall2020prediction} to our dataset for the same four-class classification. We also applied our method to a binary classification task, classifying subjects into classes of below 30 and above 30 in age. For the four-class classification task, the work in \citep{dall2020prediction} uses windows of 15 pulses which did not converge for our dataset. Instead, our data  segmentation scheme of 50 signal samples per window, was employed in place of the original 15 peaks window scheme in \citep{dall2020prediction}. For the training strategy, the learning rate at the beginning was $10^{-5}$ and decreased by a factor of 10 if the loss of the validation set did not decrease after 5 consecutive epochs. The training stopped if the validation loss did not decrease for 10 consecutive epochs. Tables \ref{table:Four-class} and \ref{table:Binary} show the results of the proposed VGTL-net method and the comparison method from Dall'Olio \textit{et al.} \citep{dall2020prediction}. Observe that the VGTL-net outperformed the comparison method in both multiclass and binary classification problems. For the multiclass classification problem, the VGTL-net converged much faster than the comparison method. Figure \ref{fig:CM4} and \ref{fig:CM2}  show the confusion matrices for the multiclass and binary classification problems of one of the train/validation/test splits.

\begin{figure}[h]
    \centering
    \includegraphics[width = 0.5\textwidth]{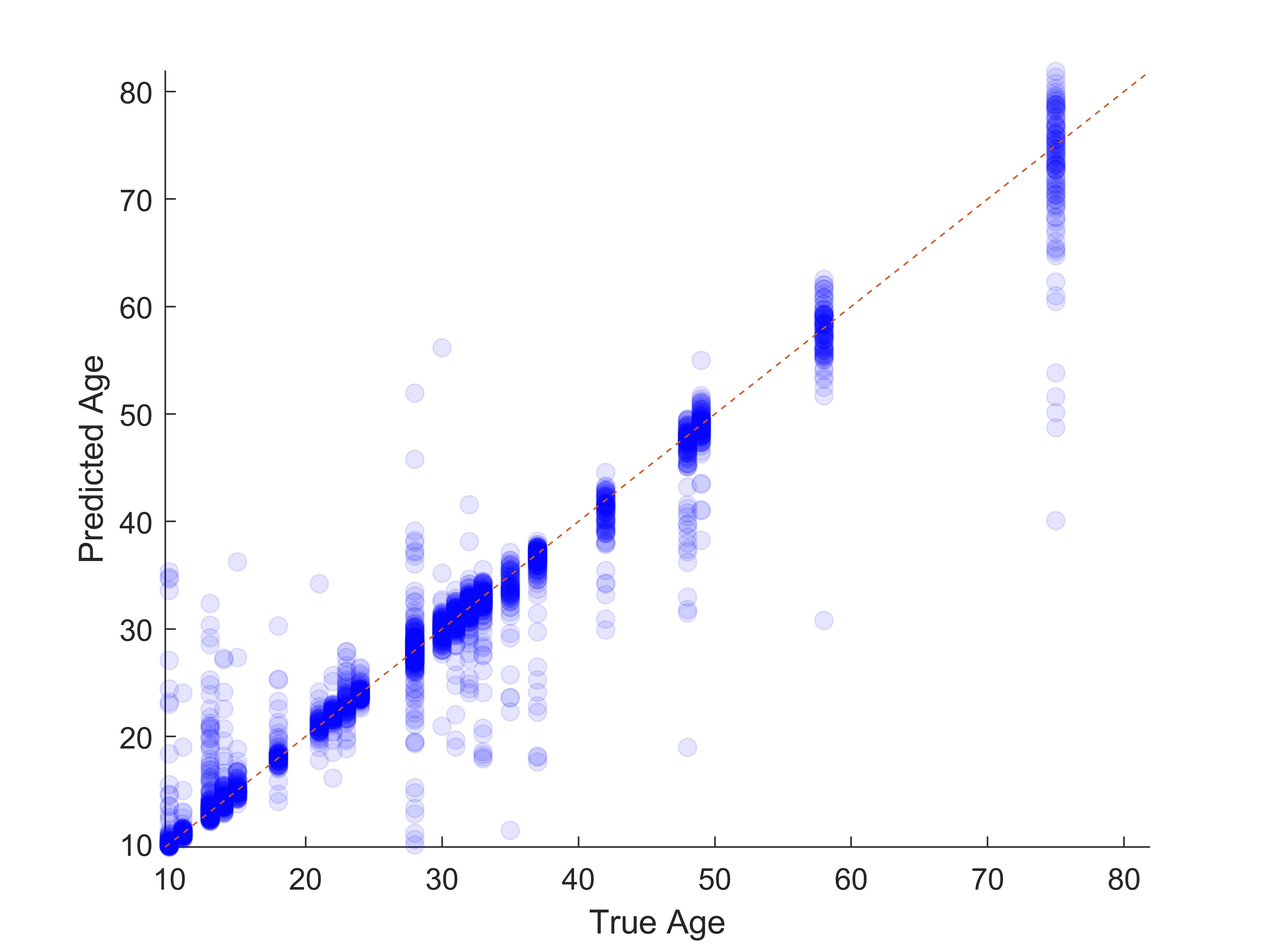}
    \caption{The results of age prediction by the VGTL-net, against the ground truth age of the subjects in the test set.}
    \label{fig:age_reg}
\end{figure}

In the regression task, the proposed VGTL-net model demonstrated high performance from the perspective of average mean absolute error and standard deviation, with the values of 1.41 years and 0.14 years, respectively. Figure \ref{fig:age_reg} shows the predicted age and ground truth age of subjects in the test set. To the best of our knowledge, this is the first study to perform regression on subjects’ age.

In the preprocessing part, the method proposed by Dall’Olio \textit{et al.} \citep{dall2020prediction} firstly removed the trend of the PPG signal using a centered moving average. The so processed PPG signal was then demodulated using a Hilbert transform and the envelope was extracted. The final version of the PPG signal was obtained by dividing the demodulated signal by the envelope. In contrast, our method did not require any preprocessing and simply used the \textbf{findpeaks} function in \textbf{MATLAB}. In addition, both methods used deep learning to classify the age group. The comparison method adapted the ResNet model from the computer vision area to work for one-dimensional signals. On the other hand, our method is naturally suitable for much more developed convolutional neural networks for two-dimensional images,  as it easily converts a 1D signal into a 2D signal, which enables us to utilise the power of transfer learning, thus resulting in faster convergence and better performance. Overall, our proposed VGTL-net showed advantages in both preprocessing and learning, as VGTL-net required much less preprocessing and did not require complex human-designed and dataset-oriented rules. 
\begin{table}[H]\centering
	\begin{tabular}{ccc} 
	\hline \text{Method} & \text{Accuracy: Four Classes} & \text{Epochs}\\
        \hline \text{VGTL-net} & \text{97.20 $\pm$ 0.2393\%} &\text{26.66 $\pm$ 3}\\
        \text{Dall'Olio, \textit{et al.}}& \text{87.14\%} &\text{280}\\
        \hline
	\end{tabular}
	\caption{Multiclass classification performances of the proposed VGTL-net and the method in \citep{dall2020prediction}.}\label{table:Four-class}
\end{table}

\begin{table}[H]\centering
	\begin{tabular}{ccc} 
	\hline \text{Method} & \text{Accuracy: Binary Classes} & \text{Epochs}\\
        \hline \text{VGTL-net} & \text{95.7 $\pm$ 0.1375\%} &\text{22 $\pm$ 1.35}\\
        \text{Dall'Olio, \textit{et al.}} & \text{95.3\%} &\text{Not Known}\\
        \hline
	\end{tabular}
	\caption{Binary classification performances of the proposed VGTL-net and the method in \citep{dall2020prediction}.}\label{table:Binary}
\end{table}

\section{Discussion}
\label{sec:conc}
We have proposed a novel model called the VGTL-net (Visibility Graph and Transfer Learning)  that combines graph theory, computer vision and deep learning to analyse photoplethysmography (PPG) signals. The method first converts the PPG signals to visibility graphs and represents them as adjacency matrices that can easily be visualised as images. This transforms the one-dimensional PPG signals into their two-dimensional representations, which are suitable as inputs to computer vision networks. We have demonstrated that visibility graphs can capture the physiological characteristics of their adjacency matrices, which underpins the VGTL-net model. Moreover, the affine transformation invariant property of the visibility graph makes it robust for PPG signals that are constantly affected by baseline wandering and motion artefacts. \textit{Given that the visibility graph only preserves the structural information and discards the amplitude information}, this is advantageous in situations where the amplitude is influenced by many exogenous factors, such as age, which typically yields misleading results. After extracting the adjacency matrices as images, we applied the pretrained models to these images to produce the final output. This enables one-dimensional PPG signals to leverage the power of computer vision algorithms that only work on two-dimensional images.

We have tested the VGTL-net model on the prediction of the subjects’ age and blood pressure from pure PPG signals, and it has shown its powerful generalisation abilities by performing classification and regression on two different tasks and two different datasets. The VGTL-net model has demonstrated its advantage of requiring fewer preprocessing steps, and the absence of any requirement to apply algorithms such as principal component analysis (PCA) to retrieve information, or empirical mode decomposition (EMD) to single out signal components. This makes the model robust and straightforward to apply since many of the preprocessing algorithms are sensitive to parameters and require human calibration.  Another advantage of VGTL-net is the elimination of the need for manually extracting features from PPG signals, as the PPG signals are fed to the VGTL-net model as a whole. This is critically important for signals of low quality that might affect feature extraction algorithms.

In summary, the VGTL-net algorithm provides a simple yet powerful framework that has good generalisation ability and has the potential to become a universal framework that is suitable for multiple applications, especially those based on physiological data

%%\section*{Acknowledgements}
%%Thanks to ...

%% The Appendices part is started with the command \appendix;
%% appendix sections are then done as normal sections
%%\appendix

%%\section{Appendix title 1}
%% \label{}

%%\section{Appendix title 2}
%% \label{}

%% If you have bibdatabase file and want bibtex to generate the
%% bibitems, please use
%%

\bibliographystyle{elsarticle-harv} 
\bibliography{ref}

%\printbibliography

%% else use the following coding to input the bibitems directly in the
%% TeX file.

%%\begin{thebibliography}{00}

%% \bibitem[Author(year)]{label}
%% For example:

%% \bibitem[Aladro et al.(2015)]{Aladro15} Aladro, R., Martín, S., Riquelme, D., et al. 2015, \aas, 579, A101

%%\end{thebibliography}

\end{document}